\documentclass[useAMS,usenatbib]{mn2e}

\def\msun{\hbox{$M_{\odot}$}~}
\def\gojko{\hbox{Djura\u{s}evi\'{c} }}

\def\rsun{$R_{\odot}$}
\def\msun{$M_{\odot}$}

\usepackage{graphicx}  
 \usepackage{times}
  \usepackage{lscape}
  \usepackage{rotating}
  \usepackage{txfonts}

\title[W\,Serpentids and Double Periodic Variables]{Interacting binaries W\,Serpentids and Double Periodic Variables}
\author[Mennickent, Otero \& Ko{\l}aczkowski]
  {R.E. Mennickent$^{1}$\thanks{E-mail: rmennick@astroudec.cl},  
     S. Otero$^{2}$, Z. Ko{\l}aczkowski$^{3}$  
\\
  $^1$Universidad de Concepci\'on, Departamento de Astronom\'{\i}a,
      Casilla 160-C, Concepci\'on, Chile\\
  $^{2}$ Buenos Aires, Argentina; American Association of Variable Star Observers (AAVSO), Cambridge, MA, USA  \\
  $^{3}$ Instytut Astronomiczny Uniwersytetu Wroclawskiego, Kopernika 11, 51-622 Wroclaw, Poland 
   }
\date{}

\begin{document}


\maketitle 

\begin{abstract} 

 W\,Serpentids and Double Periodic Variables (DPVs) are candidates for close interacting binaries in a non-conservative evolutionary stage;
while W\,Serpentids are defined by high-excitation ultraviolet emission lines present during most orbital phases, and by usually showing variable orbital periods, DPVs are characterized by
a long photometric cycle lasting roughly 33 times the (practically constant) orbital period.
We report the discovery of 7 new Galactic DPVs,  increasing the number of known DPVs in our Galaxy by 50\%. 
We find that DPVs are tangential-impact systems, i.e. their primaries  have radii  barely larger than the critical Lubow-Shu radius. These systems are expected to show transient discs, but we find that they host stable discs with  radii smaller than the tidal radius. Among tangential-impact systems including DPVs and semi-detached Algols, only DPVs  have primaries with masses between 7 and 10 $M_{\odot}$. We find that DPVs  are in a Case-B mass transfer stage with donor masses between 1 and 2 M$_{\odot}$ and with primaries  resembling Be stars. W\,Serpentids are impact and non-impact systems, 
 their discs  extend until the last non-intersecting orbit and show a larger range  of stellar mass and mass ratio than
DPVs.
Infrared photometry reveals significant color excesses in many DPVs and W\,Serpentids, usually larger for the latter ones, suggesting variable amounts of circumstellar matter. 
\end{abstract}


\section{Introduction: Unsolved problems in binary star evolution}

Since most stars in the universe are binaries or members of gravitationally bounded systems, binary star evolution constitutes a primordial subject for understanding large stellar populations.
The case of intermediate-mass binaries of the Algol type is especially interesting since epochs of severe interaction must exist to account for the mass ratio
distribution (Sarna 1993, Van Rensbergen et al. 2011, de Mink et al. 2014). In spite of the importance of the binary interaction stage, 
is not clear how efficient the processes of mass and angular momentum transfer are, remaining unknown  how much matter is 
deposited into the interstellar medium and how much is accreted by one of the stellar components. 



For the above reasons the identification and study of heavily interacting binaries is important to bring clues on the nature of mass loss mechanisms during epochs of binary interaction.
Two classes of interacting binaries (IBs) showing evidence of interaction are relevant in this sense:  the Double Periodic Variables (DPVs) and the W\,Serpentis stars. They are both members of the more general class
of Algols, close binaries usually consisting of a cool evolved star and a main sequence early type star. Early in its history, the secondary star would have been more massive, and evolved first until overfilling its Roche lobe. After fast mass exchange, the lobe-filling star became the less massive of the pair (e.g. Eggleton 2006).

Double Periodic Variables are semi-detached interacting binaries showing a long photometric periodicity lasting about 33 times the orbital period (Mennickent et al. 2003); a couple of them have been found in the Galaxy (Mennickent et al. 2012a) and more than one hundred fifty in the Magellanic Clouds (Poleski et al. 2010, Pawlak et al. 2013). An inherent feature of DPV is presence of relatively large and optically thick circumprimary disc. The long-cycle has been interpreted as evidence for cyclic mass loss (Mennickent et al. 2008), probably through a modulated disc-wind (Mennickent et al. 2012b). Surprisingly, in spite of evidence for mass loss, the orbital period remains remarkably constant in all well studied DPVs (e.g. Mennickent et al. 2012a, Mennickent 2014, Barr\'{i}a et al. 2013, Garrido et al. 2013). 

W\,Serpentis stars are interacting binaries consisting of a hot star surrounded by an optically thick accretion disc (Plavec 1980a,b, 1982, Young\& Snyder 1982).
The W\,Serpentids are characterized by strong ultraviolet emission lines of highly excited species like He\,II, C\,II, Al\,III, Fe\,III, C\,IV, Si\,IV and N\,V seen at every orbital phase, that have been thought to be formed in 
a super corona produced by the process of mass transfer and accretion (Plavec, Weiland \& Koch 1982) or by scattering in an induced stellar wind (Plavec 1989). In addition, the light curve tends to be noisy and the orbital periods variable.
The W\,Serpentids have been interpreted as semi-detached close binaries with circumprimary discs fed by Roche-lobe overflow at mass transfer  rates larger than normal Algols   (Plavec 1989). 

 The orbital eccentricity in DPVs and W\,Serpentis stars is always compatible with zero; in very few cases is very small and consistent  with the effects produced by the mass streams in the radial velocities of close binaries (Lucy 2005). This condition is in agreement with the theoretical prediction of a rapid orbit circularization for these systems by dynamical tides (Zahn 1975, 1977). 

The main goal of this paper is to compare, for the first time,  DPVs with W\,Serpentids from the observational point of view, trying to clarify if they are two separated classes of interacting binaries, and  if they are related from an evolutionary perspective.  
We also present a new census of Galactic DPVs, presenting 7 new systems, increasing significantly the number of these objects in our Galaxy.


The paper is organized as follow: in Section 2 we present our methodology for searching for new DPVs and give sources for infrared photometric data. In Section 3 the results of this search are presented, 
along with a compilation and analysis of DPV and W\,Serpentids data found in the literature. 
In this section we also analyze 2MASS and WISE colors for the sample stars. 
In Section 4 a discussion of our results is presented. We finish with our conclusions in Section 5.

\section{Methodology and data sources}

\subsection{Search for new Galactic Double Periodic Variables}

The new DPVs presented in this study were found by one of us (S.O.) as part of a multi-survey variable star search using three publicly available databases: the All Sky Automated Survey - ASAS-3 (Pojmanski 2002), the Northern Sky Variability Survey - NSVS (Wozniak et al. 2004) and the Hipparcos Catalogue (Perryman et al. 1997). More than 3000 new variable stars were found and hundreds of new/revised classifications of known variables were made. Here we report the discovery of
7 new galactic DPVs, clearly distinguished by their  blueish colors, two periods and characteristic period ratio.



\subsection{Period determination and light curve disentangling}

The photometric periods of new DPVs were determined with the IRAF PDM task (Stellingwerf 1978). Then we disentangled the two main photometric frequencies by using a code specially designed for this purpose (e.g. Mennickent et al. 2012a). The code adjusts the orbital signal with a Fourier series consisting of the fundamental frequency plus its harmonics. Then it removes this signal from the original time series letting the long periodicity present in a residual light curve. The program fits this remaining signal with another Fourier series consisting on fundamental frequency and harmonics and remove it. As result we obtain the cleaned light curve with no additional frequencies and two light curves for the isolated orbital and long periods. 

\tiny  
\begin{table*}
\centering
 \caption{Orbital and long-cycle ephemerides determined in this paper for 7 new DPVs and TYC 7398-2542-1. }
 \begin{tabular}{@{}lrcrc@{}}
 \hline
System& $P_{o}$& HJD$_{o}$& $P_{l}$& HJD$_{l}$ \\
 & (d)& (245\,0000 +) &(d)  & (245\,0000 +) \\
\hline
BF\,Cir& 6.4592 $\pm$  0.0003& 1905.1538 $\pm$ 0.0323& 219 $\pm$  2& 1846.83 $\pm$  3 \\
CZ\,Cam& 8.055  $\pm$  0.015& 7871.4039 $\pm$ 0.3222& 267 $\pm$  3& 5771.00 $\pm$  8 \\
HD\,151582& 5.823  $\pm$  0.003& 1930.6928 $\pm$ 0.1165& 160 $\pm$  2& 1779.02 $\pm$  4 \\
HD\,256413& 6.775  $\pm$  0.004& 2616.9967 $\pm$ 0.2032& 242 $\pm$  12& 2621.77 $\pm$  20 \\
V761\,Mon& 7.754 $\pm$  0.002& 1868.7884 $\pm$ 0.0775& 268 $\pm$  2& 1699.54 $\pm$  3 \\
TYC\,7398-2542-1  &2.76903 $\pm$   0.00002 &1953.8027 $\pm$  0.0055& 106 $\pm$  1& 1874.99 $\pm$  5 \\
TYC\,8627-1591-1  & 7.462 $\pm$   0.001& 1884.9720 $\pm$ 0.3731& 268 $\pm$  3& 1654.21 $\pm$  5 \\
V1001\,Cen& 6.736  $\pm$  0.003& 2441.9293 $\pm$ 0.1347& 247 $\pm$  3& 2439.69 $\pm$  5 \\
\hline
\end{tabular}
\end{table*}
\normalsize

\subsection{W\,Serpentis stars in the literature}

While DPV publications are relatively recent and we can easily find data in the literature for a statistical analysis, data for W\,Serpentids are more disperse and not so easily  accessible;  a brief description of these data is given below.

We searched for physical data of W\,Serpentids in the literature, starting with the catalog of Peculiar Emission Line Algols of Gudel \& Elias (1996). However, we found that several of their
objects proposed as W\,Serpentids are of other nature and were not included in our analysis. Here we provide a list of these objects in order to avoid confusion in future investigations.  
RZ\,Sct is proposed as a double contact binary  (Olson \& Etzel 1994), $\o$ Leo (14 Leo = HD 83808) is an evolved Am binary (Griffin 2002), V644\,Mon, KX\,And and AX\,Mon are long period Be star binaries (Halbedel 1989, Tarasov, Berdyugina \& Berdyugin 1998, Puss \& Leedjarv 1997) and SS\,Cam an eclipsing RS CVn system (Arnold et al. 1979). In addition, after searching the ADS we find that in the long-period Algols WW\,And, KU\,Cyg, RX\,Gem, DN\,Ori, AY\,Per, the Algols with cyclic period changes WW\,Cyg, SW\,Cyg and Y\,PsC, and the binaries SY\,And, UZ\,Cyg, RW\,Gem, TX\,UMa, TU\,Mon, U\,CrB, AM\,Aur and 14\,Ser, there is no reported evidence for a W\,Serpentis type  classification. Finally, AU\,Mon is a Double Periodic Variable discussed in this paper and RW\,Tau ($P_{o}$ = 2.77 days with variable period changes, Simon 1977, no physical data available) is classified as a weak W\,Serpentis star probably related to transitional objects like V356\,Sgr and U\,Cep (Plavec \& Dobias 1983).

The remaining 6 systems in the Gudel \& Elias catalog, truly classified as W\,Serpentids, were investigated for fundamental system and stellar parameters in the literature, viz., RX\,Cas, SX\,Cas, V367\,Cyg, RY\,Per, W\,Ser and RS\,Cep.
We added UX\,Mon, W\,Cru and BY\,Cru from the list of W\,Serpentids by Wilson et al. (1984) and also added the prototype of the class $\beta$ Lyrae, amounting to 10 W\,Serpentis stars.

The following stars have been mentioned as possible W\,Serpentids in the literature but they are not included in this work for the reasons given below:
V453\,Sco (HD\,163181) only shows high excitation UV emission lines at orbital phase 0.6 (Hutchings \& Van Heteren 1981),
BM\,Cas is a binary in a common envelope phase (Pustylnik, Kalv \& Harvig 2007), U\,Cep, KU\,Cyg,
RZ\,Oph and V356\,Sgr are classified as transition objects by Plavec (1989). For V1507\,Cyg (HD\, 187399) we did not find reports of ultraviolet emission
lines, whereas AG\,Peg and AR\,Pav  are symbiotic binaries (Yoo 2008, Quiroga et al. 2002, respectively). 
Finally, HD\,207739 shows variable ultraviolet emission and probably is a transition object (Kondo, McCluskey \& Parsons 1985).

\subsection{Three systems previously classified as DPV candidates}

In a recent paper, Mennickent \& Rosales (2014) report the discovery of  two new Galactic DPVs, viz.\, V495\,Cen and V4142\,Sgr, which are included in this paper. 
In addition, they classified as DPV candidates the following objects, due to the presence of two photometric periods in their ASAS light curves: TYC\,6083-192-1, TYC\,8638-2548-1 and UX\,Cnc.
However, due to their very long periods, of duration comparable with the time baseline, their small long-cycle amplitudes and very red colors, we argue for a RS\,CVn nature for these systems, where the long cycle probably represents cycles of sunspot-like activity. In this view,  the short periods reported previously for these objects are twice the true rotational periods of a single RS\,CVn star. Let's discuss briefly each of these systems.

TYC\,6083-192-1 is classified as E/RS with Coronal Activity in the ASAS Eclipsing Binaries Catalogue (Szczygiel et al. 2008) and published spectra indicate a K4V+M0V system (Parihar et al. 2009). Additionally, the SAO catalogue (Whipple 1966) indicates a spectrum K7 and it is classified as K5 in the FOCAT-S catalogue (Bystrov et al. 1994). The reported short period of 90.4 days assumed eclipses much wider than usual for RS\,CVn binaries and it could be twice the stellar rotational period of 
45.2 days.
The long cycle of around 3497 days implies a period ratio of about 77, much different from a typical DPV period ratio.

TYC\,8638-2548-1 is a ROSAT X-ray source with $J - K$ = 0.79 and $B - V$ = 1.13. The long cycle shows an asymmetrical light curve as most RS\,CVn systems. 
The reported short period of 101.3 days assumes eclipses, much wider than usual for RS\,CVn binaries, and could be twice the stellar rotational period of 50.65 days.
The long period of about 3400 days implies a period ratio of 67.1, unusual for a DPV. 
  
UX\,Cnc with $J - K$ = 0.69 and $B - V$ = 1.03, seems to be another RS\,CVn. The reported short period of 84.8 days 
assumes eclipses, much wider than usual for RS\,CVn binaries, and could be twice the stellar rotational period of 42.24 days.
The suggested long cycle of 2158 days implies a period ratio of 51.1, unusual for a DPV.

For the above reasons the three aforementioned systems were not included in our census of Galactic DPVs.

\begin{figure*}
\scalebox{1}[1]{\includegraphics[angle=0,width=16cm]{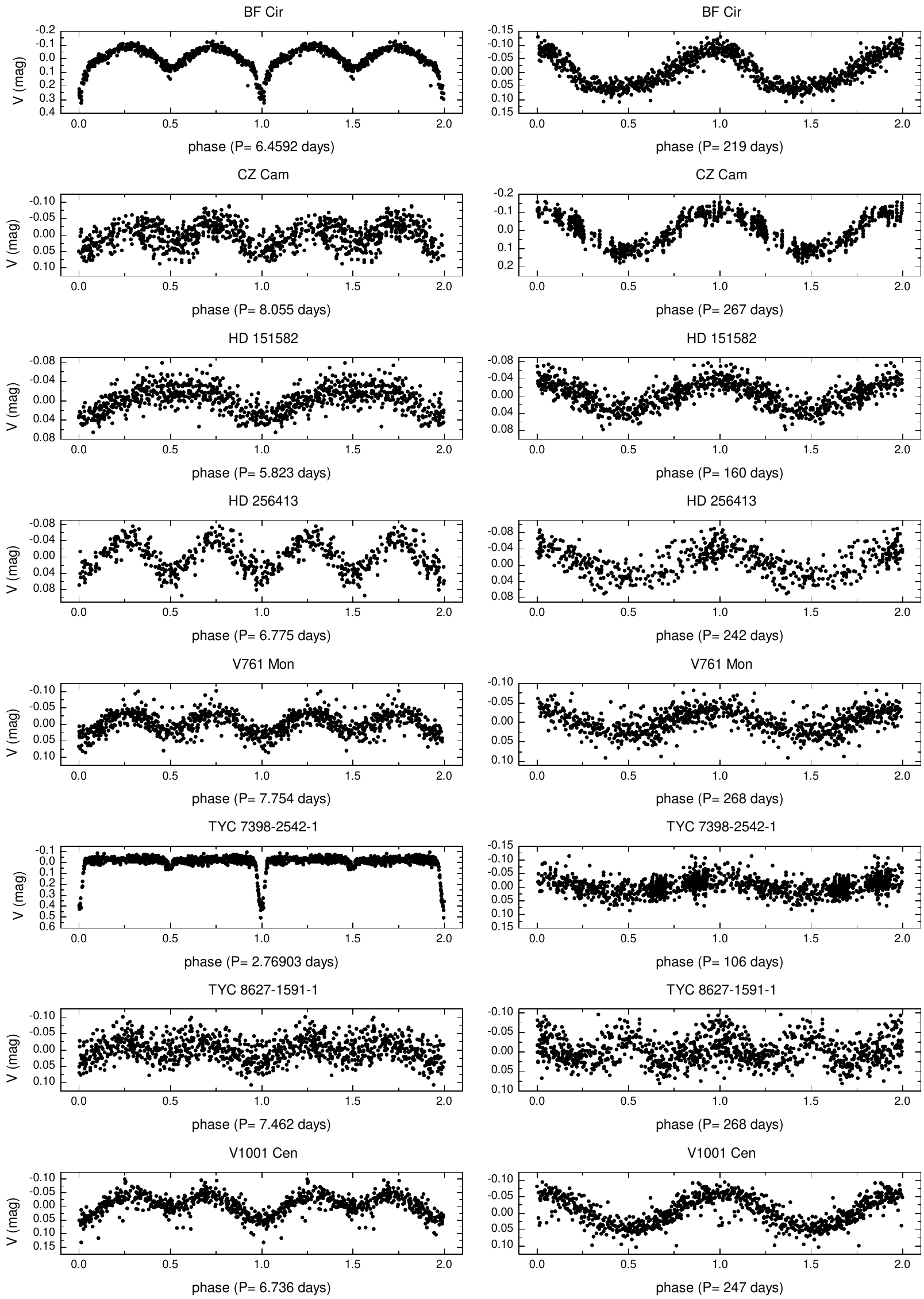}}
\caption{Disentangled orbital and long-cycle light curves for the 7 new DPVs presented in this paper and the eclipsing binary TYC 7398-2542-1 (ASAS J182841-3314.6).  They are phased according to the ephemerides given in Table 1.}
 \label{fig1}
\end{figure*}

\subsection{2MASS and WISE infrared data}

Here we give a brief summary of the infrared photometry  used in this paper. The query for magnitudes of the DPVs and W\,Serpentids  was done through the NASA/IPAC infrared science archive\footnote{http://irsa.ipac.caltech.edu/Missions/wise.html} and focused on two infrared imaging surveys.

Firstly, we scanned for data obtained with the Two Micron All Sky Survey (2MASS), a project that uniformly scanned the entire sky in three near-infrared bands, $J$ at 1.25 $\mu$m, $H$ at 1.65 $\mu$m and $Ks$  at 2.17 $\mu$m, to detect and characterize point sources brighter than about 1 mJy in each band, with signal-to-noise ratio (SNR) greater than 10, using a pixel size of 2.0" (Skrutskie et al. 2006).
The Infrared Processing and Analysis Center (IPAC) is responsible for all data processing through the Production Pipeline, and construction and distribution of the data products. Magnitude limits for point sources are at least 15.8, 15.1 and 14.3 mag at $J,H,K$ bands, respectively. 2MASS successfully obtained photometry of sources as bright as $K \sim$ 4 mag measuring the flux at the wings of the stellar image of bright stars. Unconfused point sources (with SNR $\gg$ 20) have a photometric accuracy of better than 3\%.

Secondly, we searched for magnitudes obtained with  The Wide-field Infrared Survey Explorer (WISE), a NASA medium class explorer mission that conducted
an all-sky survey at mid-infrared bandpasses centered around wavelengths 3.4, 4.6, 12 and 22 $\mu$m (hereafter
$W1, W2, W3$ and $W4$; Wright et al. 2010). 
The survey was conducted with a 40 cm cryogenically-cooled telescope in sun-synchronous polar orbit. 
Four infrared detectors imaged the same sky field of view during 7.7 s ($W1, W2$) and 8.8 s ($W3, W4$). 
We use here the data of the second-pass processing,
obtained with improved calibration and processing algorithms, superseding those obtained for the preliminary data release.  
Sources in the All-Sky Release Source Catalog saturate the WISE detectors at characteristic magnitudes of 8.1, 6.7, 3.8 and -0.4 mag in $W1, W2, W3$ and $W4$, respectively. Profile-fitting photometry  extracts reliable measurements of saturated sources using the non-saturated wings of their profiles up to brightnesses of $\approx$  2.0, 1.5, -3.0 and -4.0 mag in $W1, W2, W3$ and $W4$.

The analysis of 2MASS and WISE magnitudes is presented in Section 3.6.

\begin{figure*}
\scalebox{1}[1]{\includegraphics[angle=0,width=12cm]{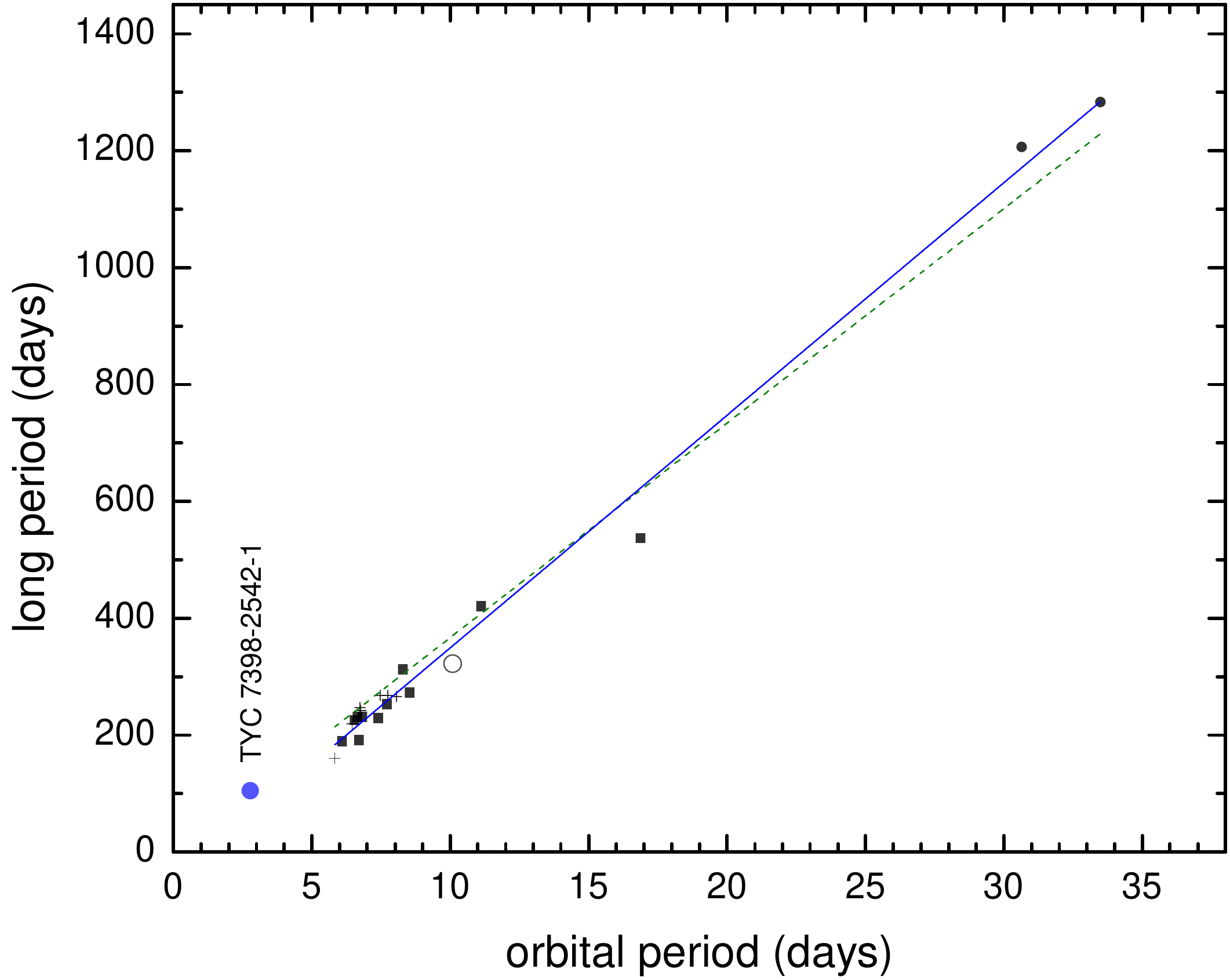}}
\caption{Orbital and long-cycle period for Galactic Double Periodic Variables and the eclipsing binary TYC 7398-2542-1. The 7 new DPVs presented in this paper (pluses) are included with those by Mennickent \& Rosales (2014; filled circles) and those by Mennickent et. al. (2012a, squares), along with V360\,Lac (Hill et al. 1997, open circle).  The best linear fits determined by least square fitting are also shown; one is forced to pass by the origin (dashed line)  and the other has the best intersection in the y-axis (solid line).}
  \label{fig2}
\end{figure*}

\subsection{About stellar and disc parameters.}

 Since our work relies on a compilation of data available in the literature, some words are necessary about the methods that have been utilized to obtain the stellar and disc parameters of the systems under study.

Parameters for DPVs are more homogeneously determined than in W\,Serpentids. In most cases the light curve was modeled with a code 
created by Djura{\v s}evi{\'c} (1992a,b). 
The code uses the inverse-problem solving method based on the simplex algorithm, and the model of a binary system with a disc. 
Spectroscopically derived data (for instance temperature of the donor or the system mass ratio) are usually used as input to constrain the light curve solution. Stellar and disc parameters and their errors are provided by the authors as the formal solution  of the aforementioned model. In particular, the stellar flattening due to rotation is included and the radius  equivalent of a spherical star of the same volume is provided. Due to the use of the same methodology  in 6 of 7 cases, we don't expect large systematic errors in DPV stellar and disc parameters.

The situation is different for W\,Serpentis stars. The method described in the paragraph above was used only for one of the W\,Serpentis stars in our sample (10\%), to determine stellar parameters, and for 3 of them to determine disc parameters (30\%). 
In all  other cases different methods were used and  hard-to-quantify systematic errors are likely present. In particular, the hiding of the primary by the disk makes difficult to detect it in the spectrum. Sometimes the radial velocities for the primary are non-existent or affected by complex absorption/emission line components. These components comes from the disc and the gas stream and can reveal unknown effects of variable optical depth. In the worst cases the temperature for the primary is estimated from the $U-B$ or $B-V$ colors, and a  main sequence stage is assumed along with a mass-radius relationship. For the above reasons, although we include the formal errors provided by the authors,  we can expect larger systematic effects for the parameters of the studied W\,Serpentis stars.

\subsection{About evolutionary models}

  We provide for most DPVs the age and the 
 mass transfer rate as found in the literature. They were usually determined by comparison of observed quantities, in particular the orbital period, stellar masses, radii and luminosities, with the predictions of the  binary evolution models of  Van Rensbergen et al.  (2008). These models include cases of strong and weak tidal interaction and also some non-conservative evolutionary tracks.
A  multi-parametric $\chi^{2}$ minimization is done between observed and predicted quantities, in order to obtain the best model for every system, as described by Mennickent et al. (2012a). Although formal errors are provided by the authors, the limited grid of models and ad-hoc assumption of mass loss in the non-conservative cases, are intrinsic  limitations of the method.

\section{Results} 

\subsection{The new Double Periodic Variables}

The result of our search for new Galactic Double Periodic Variables was the discovery of 7 new DPVs whose relevant information is given in Table\,1 and light curves shown in Fig.\,1.
Ephemerides refers to the minimum of the orbital light curve and maximum brightness of the long-cycle light curve.
During our research we found the object TYC 7398-2542-1 (ASAS J182841-3314.6) as a DPV candidate with  $P_{o}$ =  2.76903 $\pm$   0.00002 days and $P_{l}$ =  106 $\pm$  1 day. However, we reject the DPV classification since the light curve is of detached type (all others DPVs are semi-detached type) and the orbital period is quite short  (all other DPVs have $P_{o}$ longer than 5 days). The long period of this object could be due to an unresolved variable companion.  The confirmed new DPVs are: BF Cir, the only eclipsing one, HD\,151582 showing a single-wave orbital light curve and CZ\,Cam, HD\,256413, HD\,58645, TYC 8627-1591-1 and V1001\,Cen showing double-wave ellipsoidal orbital variability. All long-cycle light curves are single-hump except that of TYC 8627-1591-1 showing a double-wave modulation.

The number of currently known Galactic DPVs amounts to 21; they are presented in Table\,2, along with their orbital periods (ranging from 5.8 to 33.5 days), long periods (ranging from 160 to 1283 days), period ratios, spectral types and extreme visual magnitudes.   
From the published spectral types of DPVs, it is notable the presence of an early B-type component is most systems.

The orbital and long periods of DPVs follow an almost linear tendency (Fig.\,2). The best linear fit passing by the origin is:

\begin{equation}
P_{l} = (36.7 \pm 0.7) P_{o},
\end{equation}

\noindent  with $rms$ = 38 days. This relation is slightly different than previously reported with only 13 systems in the Galaxy, viz.\, $P_{l} = (32.7 \pm 0.9) P_{o}$ (Mennickent et al. 2012a) or 
$P_{l} = 33.13 P_{o}$ for 125 DPVs in the Large Magellanic Cloud (Poleski et al. 2010).
Excluding the two longest periods we find $P_{l} = (33.4 \pm 0.6) P_{o}$ with $rms$ = 22 days, therefore the difference comes primarily from those systems. Since equation (1)
gives oddly displaced residuals, we can get a better fit for the data with:

\begin{equation}
P_{l} = (-48.7 \pm 10.6 ) + (39.8 \pm 0.8) P_{o}, 
 \end{equation}

\noindent with $rms$ = 27 days.  The distribution of period ratios shows a mean of 33.9 with a standard deviation of 3.1 and minimum and maximum values of 27.48 days and 39.36 days (Fig.\,3).  We notice that the above relationships still remain limited by the low number of objects and likely by observational selection.


\subsection{Physical data of DPVs}


Physical parameters of all relatively well studied DPVs, 6 in our Galaxy and 1 in the LMC,
are given in Table\,3. We calculated luminosities using the published bolometric magnitudes, except for V360\,Lac, where we used average stellar radii and temperatures and the relationship $L = 4 \pi \sigma^{2} R^{2} T^{4}$. All DPV light curves have been modeled with an accretion disc around the more massive component, except those of V360\,Lac and LP\,Ara. The fact that
no disc was assumed in the V360\,Lac configuration could produce a bias in the system parameters and mimic a near contact configuration (Linnell et al. 2006). On the other hand, LP\,Ara light curve has been modeled with a Wilson-Devinney code constraining the light contribution of a possible accretion disc to less than 5\% of the total orbital light (Mennickent et al. 2011).  It is then very likely that all DPVs harbor accretion discs. The evolutionary stage of some DPVs has been found comparing  the system parameters with those of a grid of published evolutionary tracks allowing to obtain the system age, mass transfer rate and core hydrogen concentration for the cool and hot star of the binary pair ($X_{c}$ and $X_{h}$ in Table\,3).

An inspection of Table\,3 reveals  that whereas donors of DPVs span a considerable range of effective temperatures and spectral types, between 5.7kK to 12.9kK, the gainers cluster around early B-type stars ($T_{eff}$ between 15.9kK and 25.1kK),  consistent with the spectral classifications given in Table\,2.
The range of binary total mass for the DPV phenomenon is 8$-$13 $M_{\odot}$.
We find that all systems have reversed their mass ratio and  have $q$ $\leq$ 0.3. 

It is notable that all DPVs are practically in a full Case-B  mass transfer state with almost zero hydrogen in the donor core ($X_{c} \approx$ 0). Consequently, the donor is evolved, filling  its Roche lobe, and all systems with determined ages have been found inside or after a mass transfer rate event (see references in Table\,3). The evolutionary stage inferred from the models, inside or slightly after a mass transfer event,  is compatible with the observational evidence of circumstellar matter and accretion discs in DPVs.

Another interesting aspect of Table\,3 is that according to the listed references, the best models are found at a conservative stage.  
If the DPV phenomenon is due to cyclic mass loss as proposed by Mennickent et al. (2008), then 
this conservative character of the models could indicate that  the departure from the conservative case is probably subtle, i.e.  
the mass loss is minor compared with the accreted mass  in these systems.  Alternatively, it could mean that the system has entered the non-conservative phase just recently, something already envisaged for AU\,Mon (Mennickent 2014).
Moreover, an earlier liberal age of large mass transfer with substantial mass loss is probably discarded because of the good match with conservative (or slightly non-conservative) models. 

 
\begin{figure}
\scalebox{1}[1]{\includegraphics[angle=0,width=8cm]{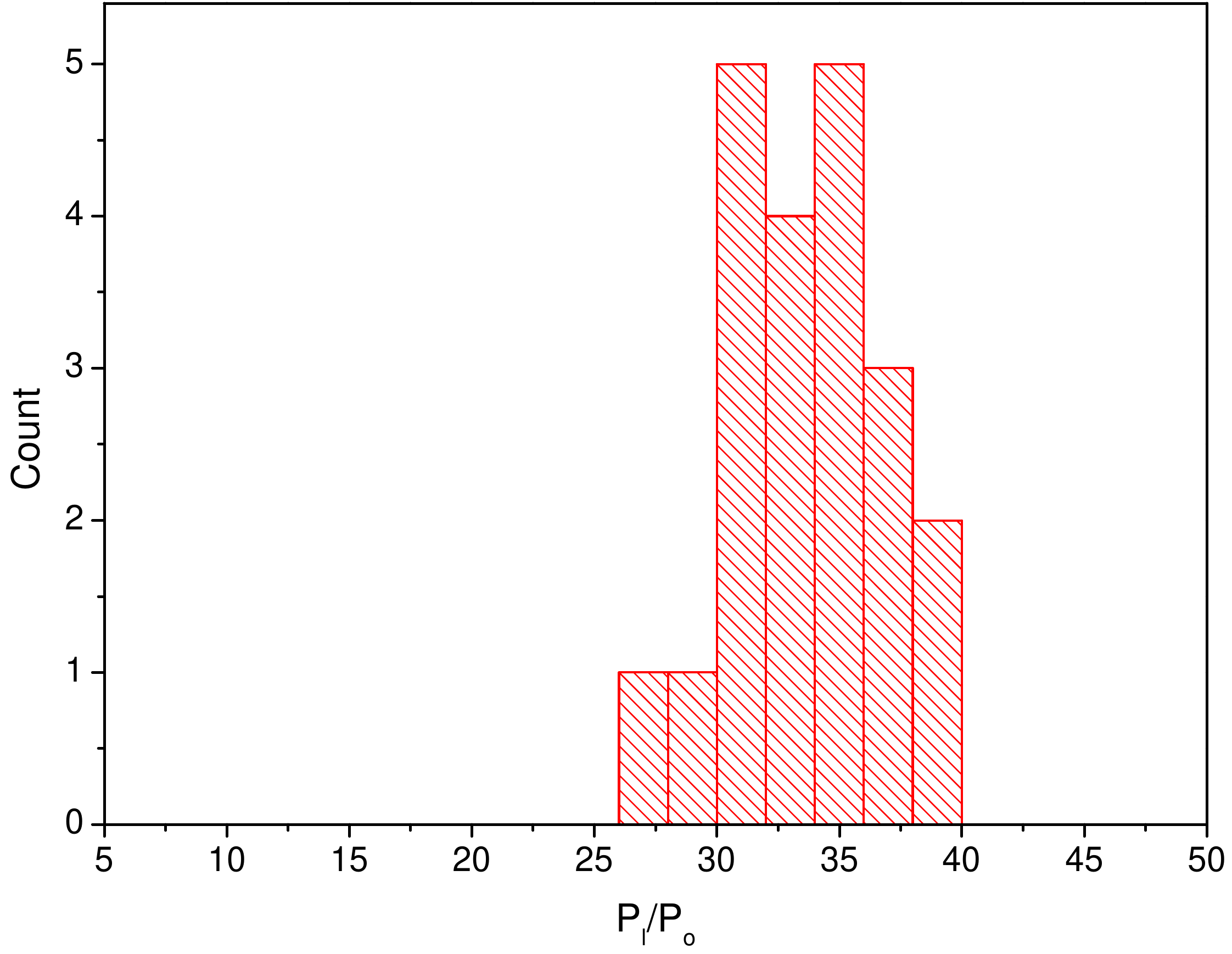}}
\caption{Histogram of period ratios for Galactic DPVs. }
  \label{fig3}
\end{figure}

\subsection{Physical data of W\,Serpentids}

We present in Tables\,4 and 5 the physical data of W\,Serpentis stars. The luminosities are calculated from the bolometric magnitudes, except for BY\,Cru, W\,Cru and RS\,Cep where we used average stellar radii and temperatures provided by the authors and the relationship $L = 4 \pi \sigma^{2} R^{2} T^{4}$. 
We find that W\,Serpentids are characterized by secondary star masses between 0.4 and 3.9 \msun, and primary star masses between 2.8 and 16.2 \msun. Total mass in W\,Serpentids, between 2.5 and 16.2 \msun, can be smaller than for DPVs. The mass ratios of W\,Serpentids span a larger range than DPVs, from 0.15 to 1.2, they are lower than 1.0 except for UX\,Mon (1.15). 
The temperatures of the secondary also have larger range than DPVs, from 4kK to 13kK.  Contrary to DPVs, W\,Ser are characterized by changing orbital periods (it increases or decreases a couple of seconds per year) and higher mass transfer rates of the order of 10$^{-5}$ to 10$^{-8}$ \msun  yr$^{-1}$. These figures are usually obtained from the rate of orbital period change, assuming conservative evolution.  The orbital periods run from 5.9 days (UX\,Mon) to 
198.5 days (W\,Cru). In our sample of DPVs and W Serpentids, the longer orbital periods are found in W\,Serpentis stars.

 Contrary to previous reports (Sudar et al. 2011), we find a constant orbital period in UX\,Mon.
The epoch of minimum published in 1950 (HJD 2433328.853; Kreiner and Ziolkowski 1978) was still valid in 2009 (epoch of ASAS-3 data) after 59 years and the best period fitting the ASAS $V$-band light curve is  5.90442 $\pm$ 0.00005 d.
The previous reports of period changes might have been influenced by  measurements of eclipse timings in a light curve of variable shape.
Disentangling the light curve in an orbital and non-orbital component, we find short-term and non-periodic residual variability with amplitude of about 0.2 mag (Fig.\,4). Since this variability decreases during primary minimum and just before secondary eclipse, it likely arises from structures in the orbital plane and inside the binary, like the accretion disc and hotspot.

\begin{figure}
\scalebox{1}[1]{\includegraphics[angle=0,width=8cm]{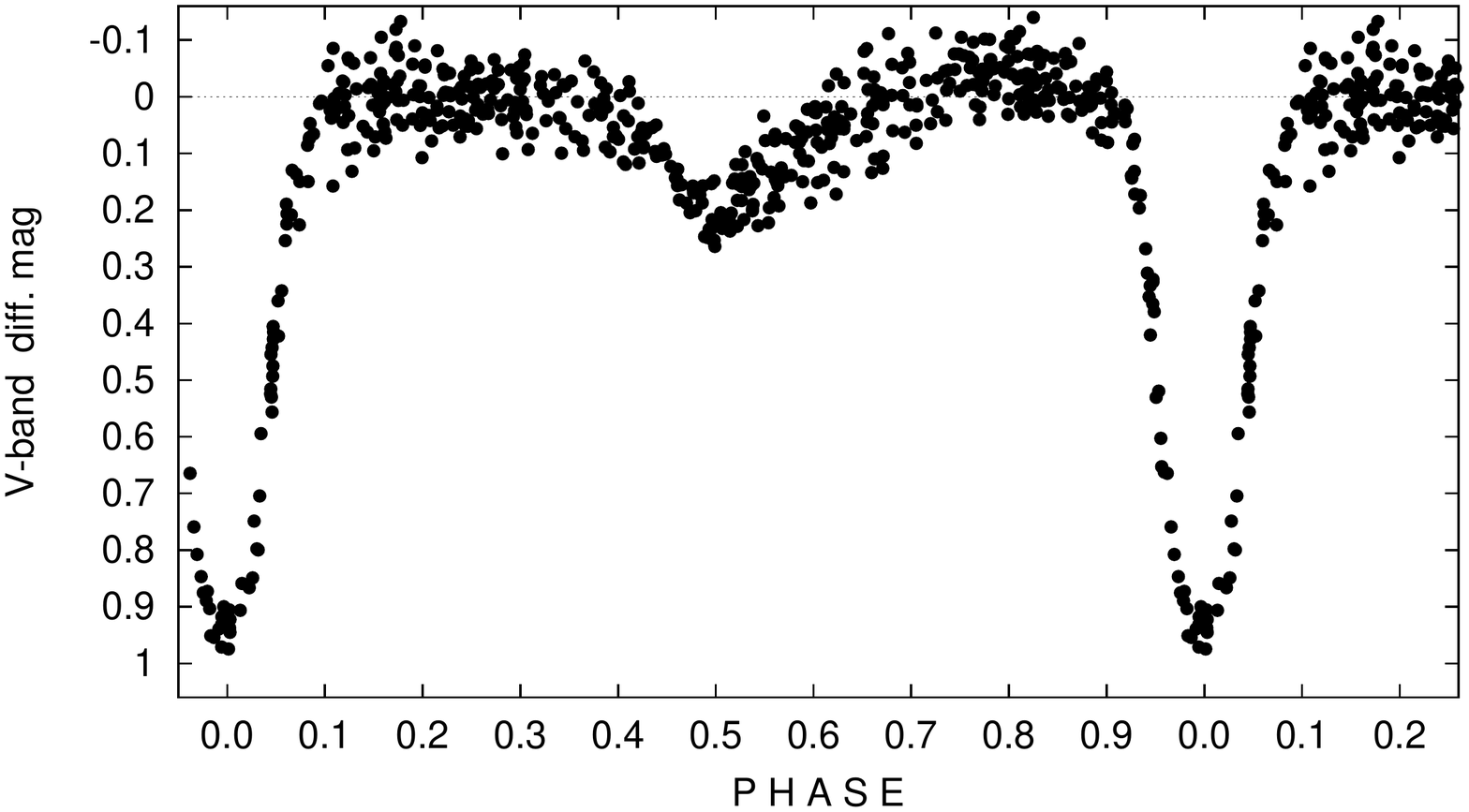}}
\scalebox{1}[1]{\includegraphics[angle=0,width=8cm]{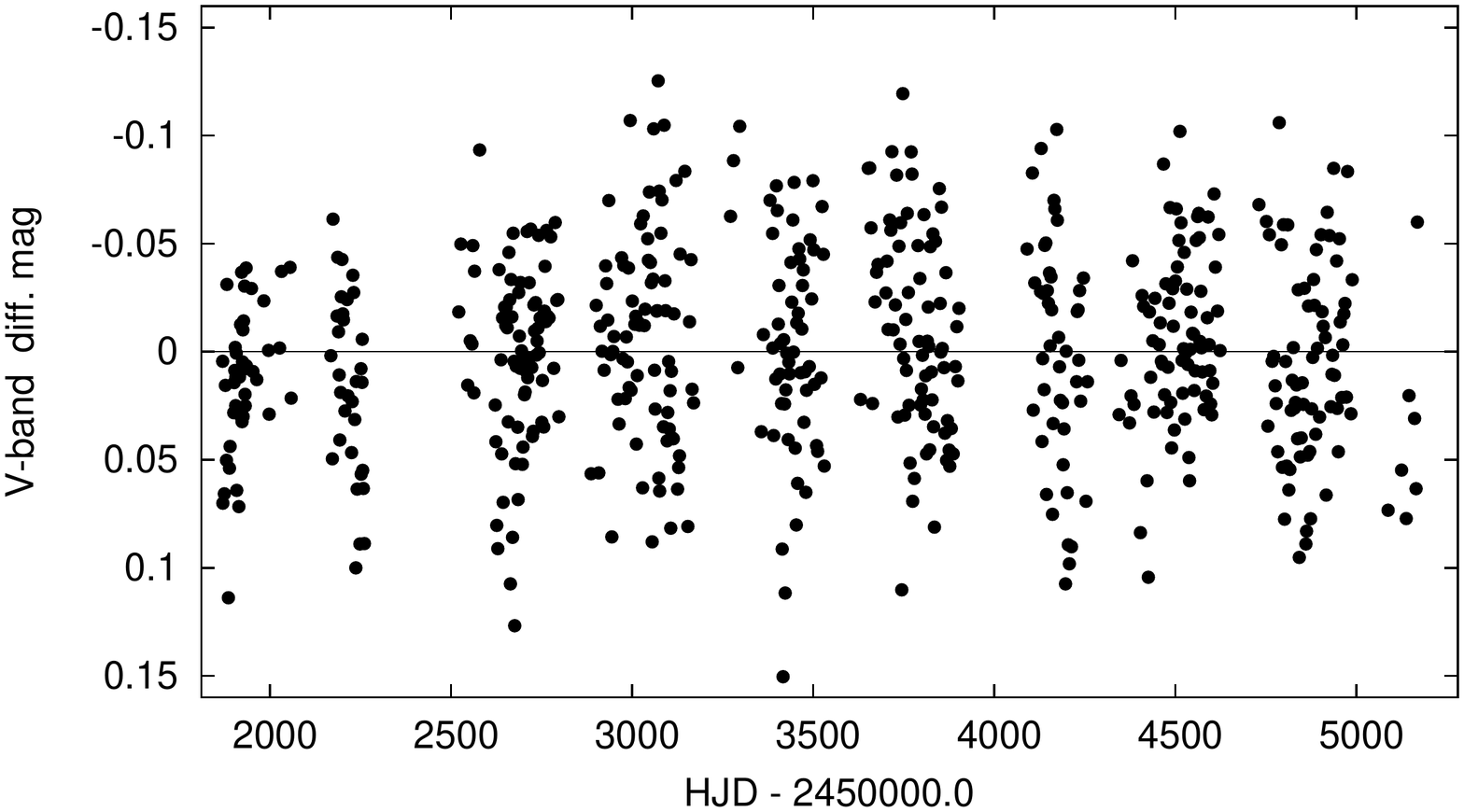}}
\caption{(up): Disentangled ASAS light curve for UX\,Mon fitted with a single period of  5.90442 d. (below): residual  light curve showing additional short-term  variability with total amplitude of about 0.2 mag. }
 \label{fig4}
\end{figure}



\subsection{DPVs and W\,Serpentids are different classes of objects}

Apart from the physical parameters discussed in the previous sections, suggesting significant differences between W\,Serpentids and DPVs,
there are also differences from the observational point of view. 

For instance, none W\,Serpentids has been found with a long photometric cycle
matching the relationship given by Eq.\,1. There are long cycles in some W\,Serpentids;  in $\beta$ Lyr a period of 282.4 d is reported (Harmanec et al. 1996) and in RX\,Cas the long cycle lasts 516.1 days (Kalv 1979).
For the $\beta$ Lyr the period ratio is 21.8 and for RX\,Cas 16.0, they do not fit the period-period relationship of DPVs. 


A search in the Mikulski Archive for Space Telescopes (MAST\footnote{https://archive.stsci.edu/index.html}) revealed that only 3 of the DPVs presented in this paper (V393\,Sco, AU\,Mon and V360\,Lac) have IUE\footnote{http://science.nasa.gov/missions/iue/}  ultraviolet spectra. Searching the existing literature we confirmed that they do not fit the
W\,Serpentid definition of showing UV emission lines at all orbital phases, suggesting  furthermore that DPVs and W\,Serpentids are two distinct types of objects.

\begin{figure}
\scalebox{1}[1]{\includegraphics[angle=0,width=8cm]{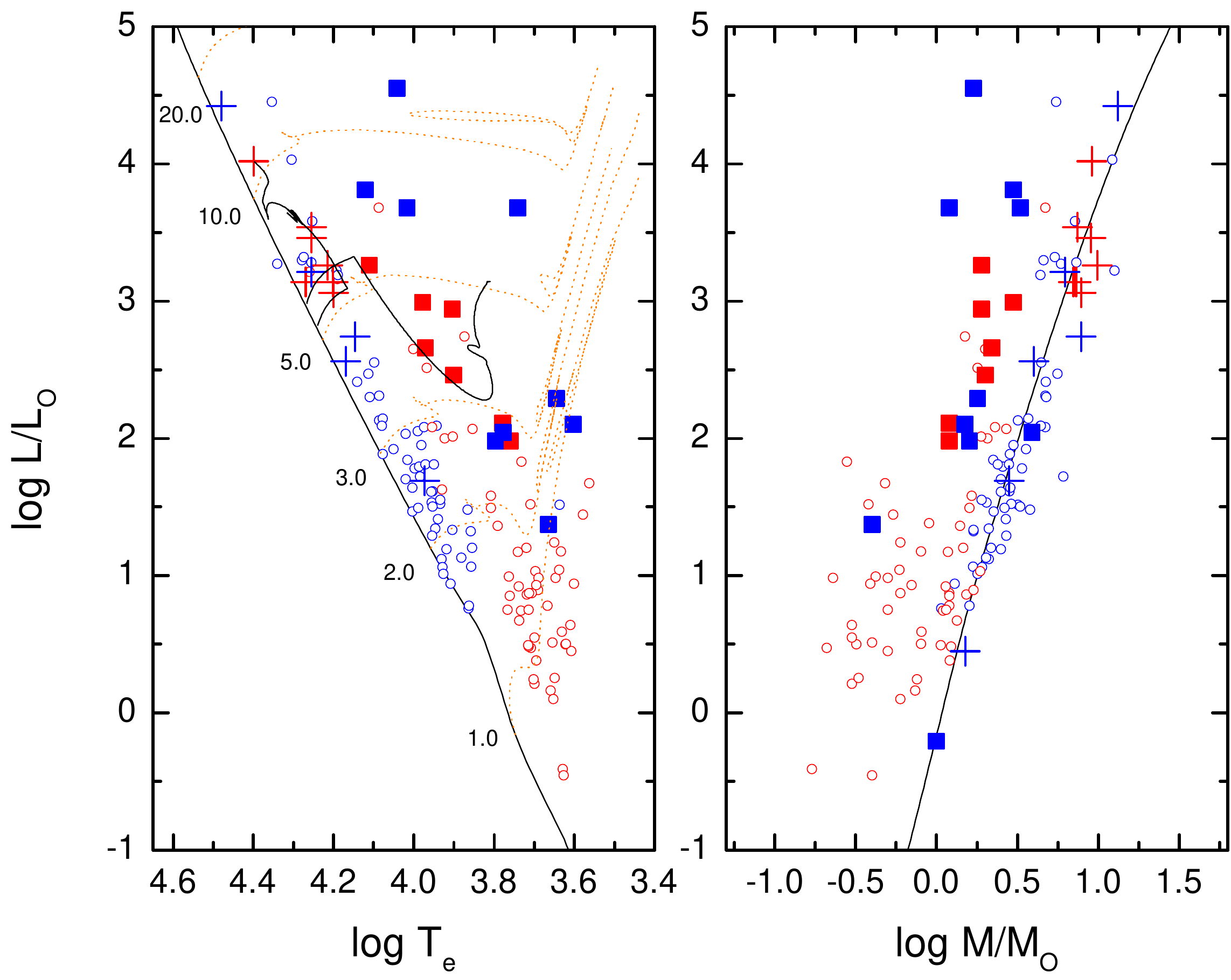}}
\caption{Comparison of physical data for semi-detached Algols from Dervi{\c s}o{\v g}lu, Tout \& Ibano{\v g}lu (2010, primaries open blue circles and secondaries open red circles), DPVs (primaries red crosses and secondaries red squares) and W\,Serpentids (primaries blue crosses and secondaries blue squares). The zero-age main sequence for $Z$ = 0.02 is plotted with a solid black line and evolutionary tracks for single stars with initial masses (in solar masses) labeled at the track footprints are also shown  (Pols et al. 1998).  The best evolutionary tracks for the primary and secondary of HD\,170582 are also plotted by two solid lines in the left panel  (see text for details).}
 \label{fig5}
\end{figure}

\subsection{Comparison with semi-detached Algols}

We compare luminosities, masses and temperatures of primaries and secondaries of W\,Serpentids, DPVs and semi-detached Algols (Fig.\,5), using as reference the loci for the main sequence for $Z$ = 0.02 and some
evolutionary tracks for single stars from Pols et al. (1998).   These are included to illustrate the degree of donor evolution, not to represent their evolutionary track, since it differs for a member of a mass-transferring binary, as illustrated by the 
binary track in the same panel (see Section 4.2).
It is clear that DPVs and W\,Serpentids  generally  possess hotter, more massive and more luminous stellar components than semi-detached Algols. In addition, 
primaries are slightly evolved, slightly displaced from the main sequence in the Log\,$L$ vs. Log\,$T_{eff}$ diagram, as occurs in Algols but secondaries are quite evolved and do not follow
the mass-radius relationship as occurs for primaries. This has been traditionally interpreted as the result of the evolution of a donor that has transferred part of its atmosphere onto the gainer resulting in a much evolved cooler star whereas the gainer remains near the main sequence (e.g. Dervi{\c s}o{\v g}lu, Tout \& Ibano{\v g}lu 2010).
It seems  that some donors of W\,Serpentids are slightly more evolved than DPV donors. When doing this comparison, we should keep in mind that catalogues of Algols might still contain some DPVs or W\,Serpentids.
For instance, Budding et al. (2004) classify as Algol 2 of our 5 Galactic DPVs and 5 of the 10 W\,Serpentids mentioned in this paper. 

 Another aspect of Fig.\,5 is the gap in donor luminosity between W\,Serpentids and DPVs; these latter occupy an intermediate region between faint and bright W\,Serpentid donors. Due to the limited dataset, this tendency could be an artifact.

\subsection{Infrared excess and circumstellar matter}

 \subsubsection{2MASS photometry of DPVs and W\,Serpentids}

2MASS colors were obtained from SIMBAD for our DPVs and W\,Serpentids. A sample of Be stars with measured $JHK$ colors was included as comparison; data are from Howells et al. (2001) who present infrared photometry of 52 isolated Be stars of spectral types O9-B9 and luminosity
classes III-V.  The purpose of including Be stars is that they are fast B-type stars surrounded by ejected disc-like circumstellar envelopes, hence  
 they provide a reference for  evaluating the effects of circumstellar material in the infrared colors of our sample stars. We also included in our study theoretical colors for dwarfs and giants obtained from Bessell et al. (1998). We did not apply interstellar extinction corrections since for most objects there is no distance estimated. However, interstellar extinction is small in infrared wavelengths and it is expected to have a small contribution to the color excess compared with circumstellar reddening.


The infrared colors of the studied stars 
indicate that all these stars show color excess in comparison with the main sequence and giant stars, and that W\,Serpentis stars in general show redder color than DPVs (Fig.\,6). 
Interestingly, Be stars occupy the same color range as DPVs. Since
Be stars show color excess mostly attributed to circumstellar reddening and proportional
to Balmer emission line strength (Howells et al. 2001, Dachs et al. 1988),  it is reasonable to assume that the
color excess observed in DPVs is also a signature of circumstellar matter, the same for W\,Serpentids showing even larger color excess.

We also find that longer orbital period systems tend to show redder $H-K$ colors  (Fig.\,7). This might be related to the presence of 
bigger and more luminous secondaries in longer period systems, but also to larger accretion discs inside the large Roche lobes of the primaries.

\begin{figure}
\scalebox{1}[1]{\includegraphics[angle=0,width=8cm]{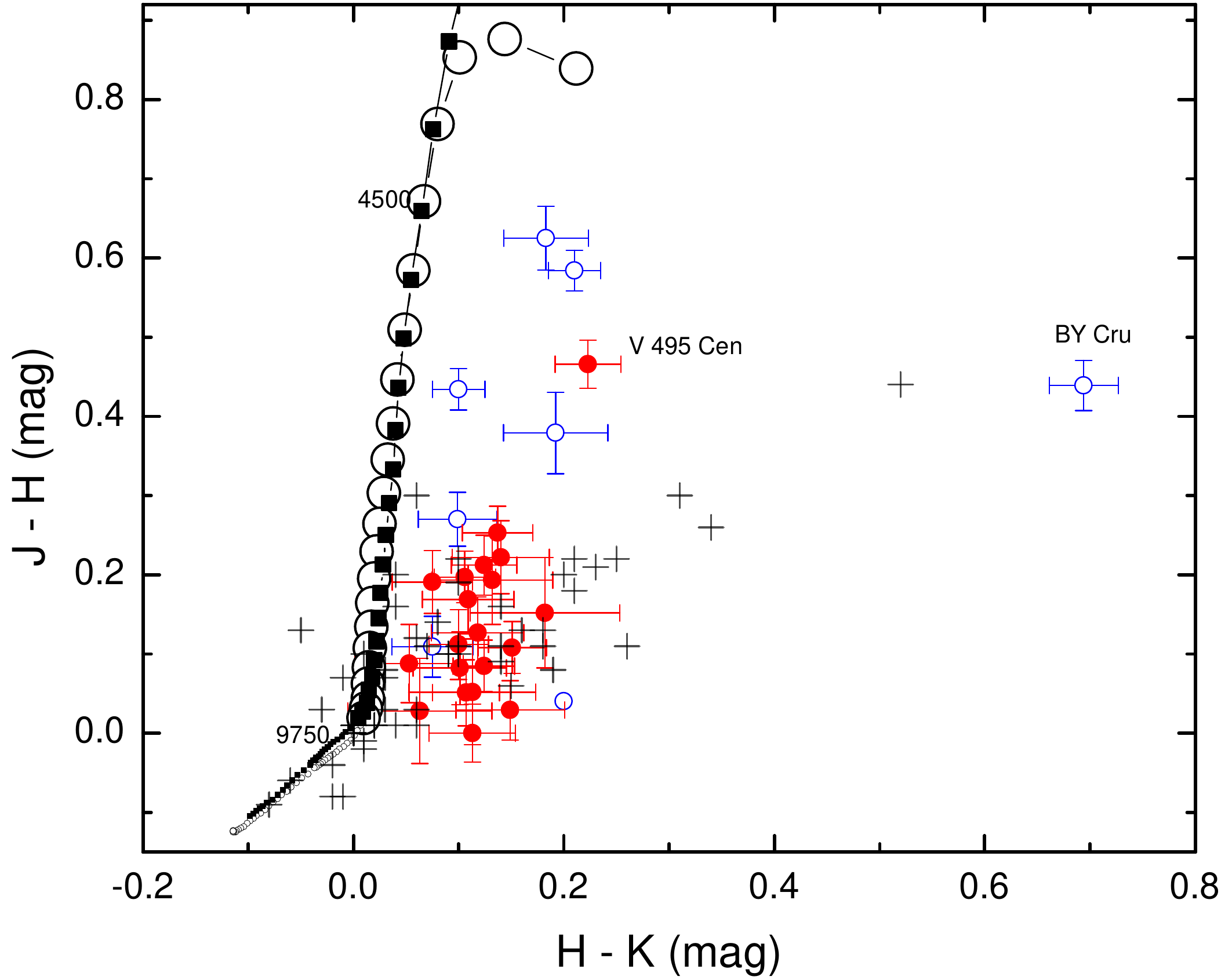}}
\caption{Color-color diagram for Galactic DPVs (red filled circles), W\,Ser stars (blue open circles), Be stars (crosses, data from Howells et al. 2001) and synthetic stellar atmosphere models with log\,g = 4.0 (squares with lines) and log\,g = 3.0 (open circles with lines, from Bessell et al. 1998). The effective temperature of 2 selected synthetic models have been labeled. }
 \label{fig6}
\end{figure}

\begin{figure}
\scalebox{1}[1]{\includegraphics[angle=0,width=8cm]{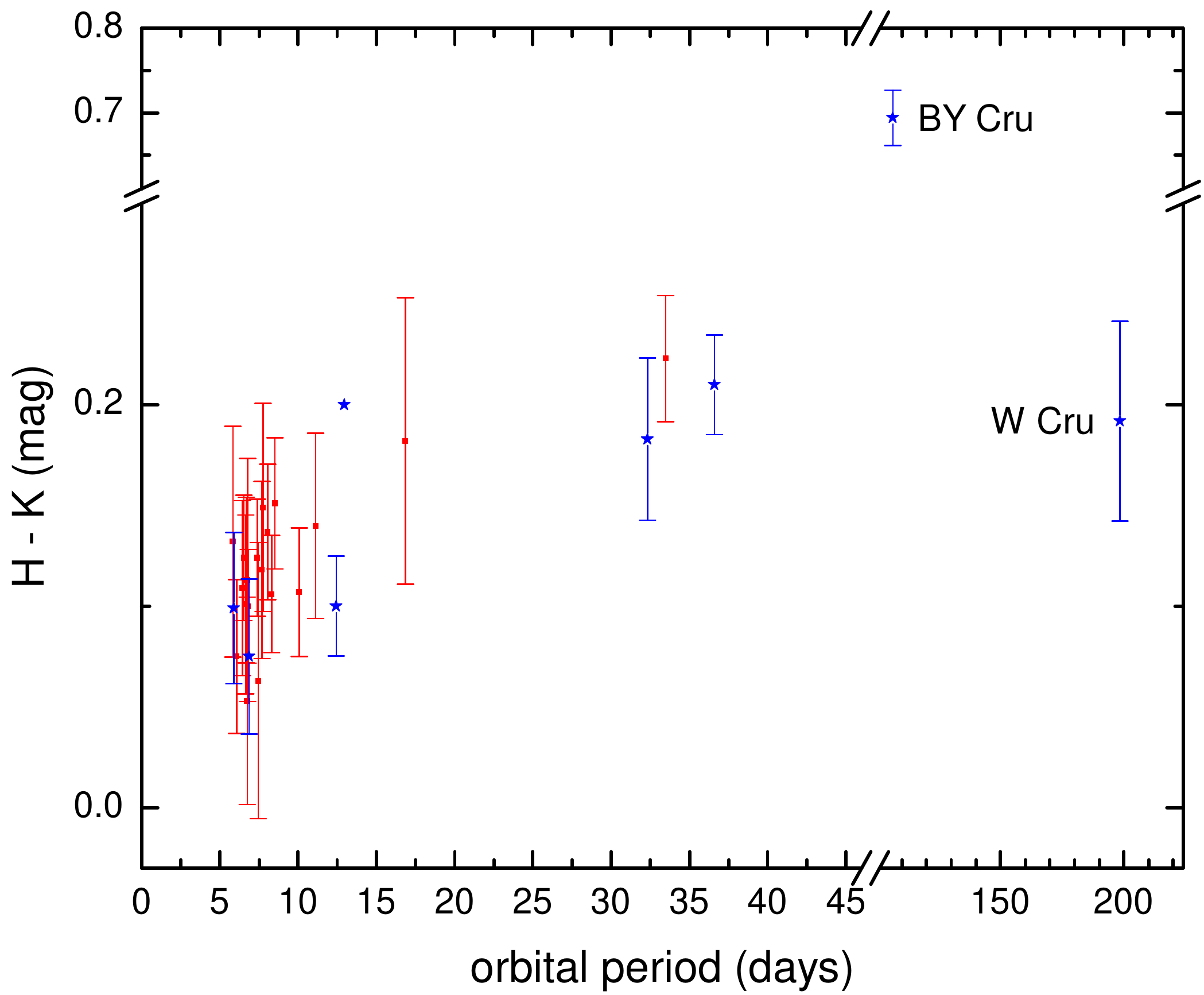}}
\caption{2MASS $H - K$ color versus orbital period for DPVs (red squares) and W\,Serpentids (blue stars).}
 \label{fig7}
\end{figure}

\subsubsection{WISE photometry of DPVs and W\,Serpentids}

We visually inspected the WISE images at bands $W1, W2, W3$ and $W4$ for all  systems. In general, all objects were detected at all bands, except some of them at $W3$ (GK\,Nor and LP\,Ara) and $W4$ (DQ\,Vel, GK\,Nor, HD\,170582, HD\,90834, LP\,Ara and TYC\,8627-1591-1).
Presence of nebulosity was not observed.

In order to include most of the objects in our study, we used only the three first WISE magnitudes to construct $W1-W2$ and $W2-W3$ color indexes. Data obtained during eclipses were discarded.  When possible, and just for few cases, we considered independently  separated epochs for a given star. In this process some additional epochs were excluded for a given star, due to the lack of simultaneous multi-band data.

We calculated the mean magnitudes for each filter ($i$= 1,2,3):\\

\begin{equation}
W_{i} = \frac{1}{n} \Sigma_{j=1}^{n} w_{i,j},
\end{equation}

\noindent
and the variance of the mean:\\

\begin{equation}
eW_{i}=     \frac{1}{\sqrt{n(n-1)}}  \sqrt{ \Sigma_{j=1}^{n} (w_{i,j}-W_{i})^{2}},
\end{equation}

\noindent
where mean magnitudes $W_{i}$ are defined for samples of $n$ individual $w_{i,j}$ magnitudes. 
These values are given for the sample stars in Tables\,6 to 9. 
We notice that  the
true variability (measured by the root mean square), can be several times the quoted variance of the mean. As in the case of $JHK$ photometry, no correction by interstellar reddening was performed.
However, at these wavelengths, the effect of interstellar reddening is neglectable.

\begin{figure}
\scalebox{1}[1]{\includegraphics[angle=0,width=11cm]{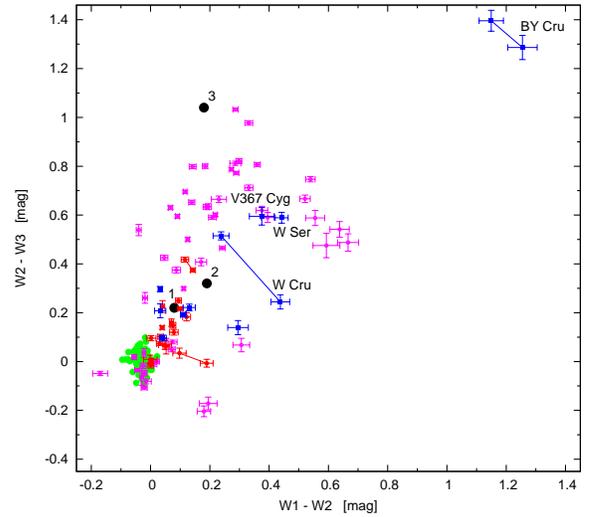}}
\caption{WISE colors for single and non-variable HIPPARCOS stars spanning the spectral-type range B1\,V to K3\,V (green dots), DPVs (red dots), Be stars (magenta circles) and W\,Serpentis stars (blue squares). 
Different epochs for a given star are represented by points joined by a line. The error bars represent variance of the mean; the observational amplitude of variability can be several times larger. The big black dots labeled with numbers refers to the models discussed in the text.}
 \label{fig8}
\end{figure}

We compared data of DPVs, W\,Serpentis stars and Be stars; the last were used as testers of circumstellar matter. 
In addition, we included as a reference  the average colors for 136
main sequence stars from the Hipparcos catalogue.  For the selection of this sample we applied several criteria:
V magnitude in range of 8 to 10, Hipparcos  magnitude ($H_{p}$) scatter less than 0.04 mag, known variables rejected,
only stars with confirmed spectral class V, all known emission line objects were rejected, all known multiple systems were discarded,
all stars with very close companions were rejected, finally few outliers were discarded. The final sample consisted of 
 stars in the range of spectral types from B1\,V to K3\,V (Tables 9 to 11).

From the distribution of systems in the color-color diagram we conclude (Fig.\,8): 
(i) The region occupied by the Hipparcos standards in the range B3\,V to K3\,V is very compact, around the origin; it means that for binary systems (without IR excess) we can
expect exactly the same colors, 
(ii) DPVs, Be stars and  W\,Serpentis stars show in general significant color excess compared to main sequence stars, (iii) 
DPVs show in general less color excess than W\,Serpentids and Be stars, (iv) some DPVs or W\,Serpentis stars show large color variability and  (v) BY\,Cru stands out showing the largest color excess in our sample (this is also true for the 2MASS $H-K$ colors).

In order to properly calibrate the color-color diagram with a real physical situation of a close binary similar to the systems considered in this study, 
we introduce some of the models calculated by Deschamps et al. (2015) for a binary loosing matter through a radiative wind formed at the stream-star
impact region. Deschamps et al. (2015) use a state of the art code to model the binary parameters during the evolution of the mass transfer episode, 
focusing on the impact of the outflowing gas and possible presence of dust grains on the spectral energy distribution. 
Models labeled 1, 2 and 3 in Fig.\,8 correspond to   models \rm{A-df-0.71},  \rm{B-df-0.71}, and \rm{B-ld-0.71}. These are the models  ``dust-free'' (df) and ``low-dust '' (ld)
at 20\,000 yr (model A) and 60\,000 years (models B) after peak of mass transfer and with  71\% of fraction of the spherical domain covered by the gas outflow (see model details in Deschamps et al. 2015). 
The models are constructed for a binary consisting of a late G-type giant and a dwarf B-type gainer located at 300 pc.  

We interpret the good match of the models 1, 2 and 3 with the position of some DPVs and W\,Ser stars as evidence of circumstellar matter in these stars.
None of the Deschamps et al. (2015) models can reproduce $W1-W2$ colors larger than 0.3 mag, neither the $J-H$ vs. $H-K$ color distribution of our sample stars; in general the predicted colors are too blue at $J-H$. The incapacity of the models in reproducing these regions of the color space could indicate still unexplored model parameters and binary configurations; the complexity of the calculations carried out by Deschamps et al. probably forced the restriction of many parameters to very particular cases. In spite of that, the conclusion that the large color excesses observed in DPVs and especially in W\,Serpentids can be interpreted as signatures of circumstellar matter seems to be well justified.

We notice that the spectral energy distribution of V393\,Sco  and HD\,170582 in the ultraviolet, optical and infrared ranges has been modeled to high accuracy with the contribution of two stellar components plus interstellar reddening, without necessity 
of introducing a circumstellar component (Mennickent et al. 2010, Mennickent et al. 2015).  This is consistent with the location of both stars in the lower left part of the color-color diagram near the main sequence region; for V393\,Sco $W_{1}-W_{2}$ = 0.002 and $W_{2}-W_{3}$ = 0.0955 and for HD\,170582 $W_{1}-W_{2}$ = 0.0786 and $W_{2}-W_{3}$ = 0.12. 

Finally we notice that longer orbital period systems tend to show redder $W2-W3$ colors (Fig.\,9). If color excess is a measure of circumstellar matter, as suggested in previous paragraphs, this 
tendency could reflect the larger Roche lobes and possibly larger discs that these systems can host.

\section{Discussion}

\subsection{Study of discs in DPVs and W\,Serpentids}

We have compiled  disc external radii and gainer radii for DPVs and W\,Serpentids from the literature (Table\,12). 
These radii relative to the binary separation are shown in Fig.\,10. 
The diagram $R_{1}/a$ vs. $q$ has been previously used to separate Algols with discs and without discs (e.g. Plavec 1989, Peters 2001). Following these authors, we included three theoretical radii in the diagram.

The first one is the 
Lubow \& Shu (1975) critical radius which can be approximated by (Hessman \& Hopp 1990):

\begin{equation}
\frac{r_{c}}{a} = 0.0859q^{-0.426}  ~~~~~\rm{for ~~0.05 < q < 1}.
\end{equation} 

\noindent
which is accurate to 1\%.
This radius is usually taken as the maximum possible radius of the primary allowing disc formation. A particle orbiting at this radius
has the same specific angular momentum as a particle released at the inner Lagrangian point $L_{1}$;  therefore, the radius corresponds to the radial extension of a ring of matter formed by mass loss due to Roche lobe overflow, just before viscosity starts spreading it into a disc. Therefore, a primary whose radius is larger than the critical radius ($R_{1} > r_{c}$) is an impact-system, where the gas stream hits the star and disc formation is unlikely. 
Actually, due to the  finite stream size, still it is possible that the outer stream orbits avoid the impact, the respective radius $r_{max}$  is a bit larger than $r_{c}$ and also a function of the mass ratio  (Lubow \& Shu 1975). This radius is also shown in Fig.\,10.

\begin{figure}
\scalebox{1}[1]{\includegraphics[angle=0,width=8cm]{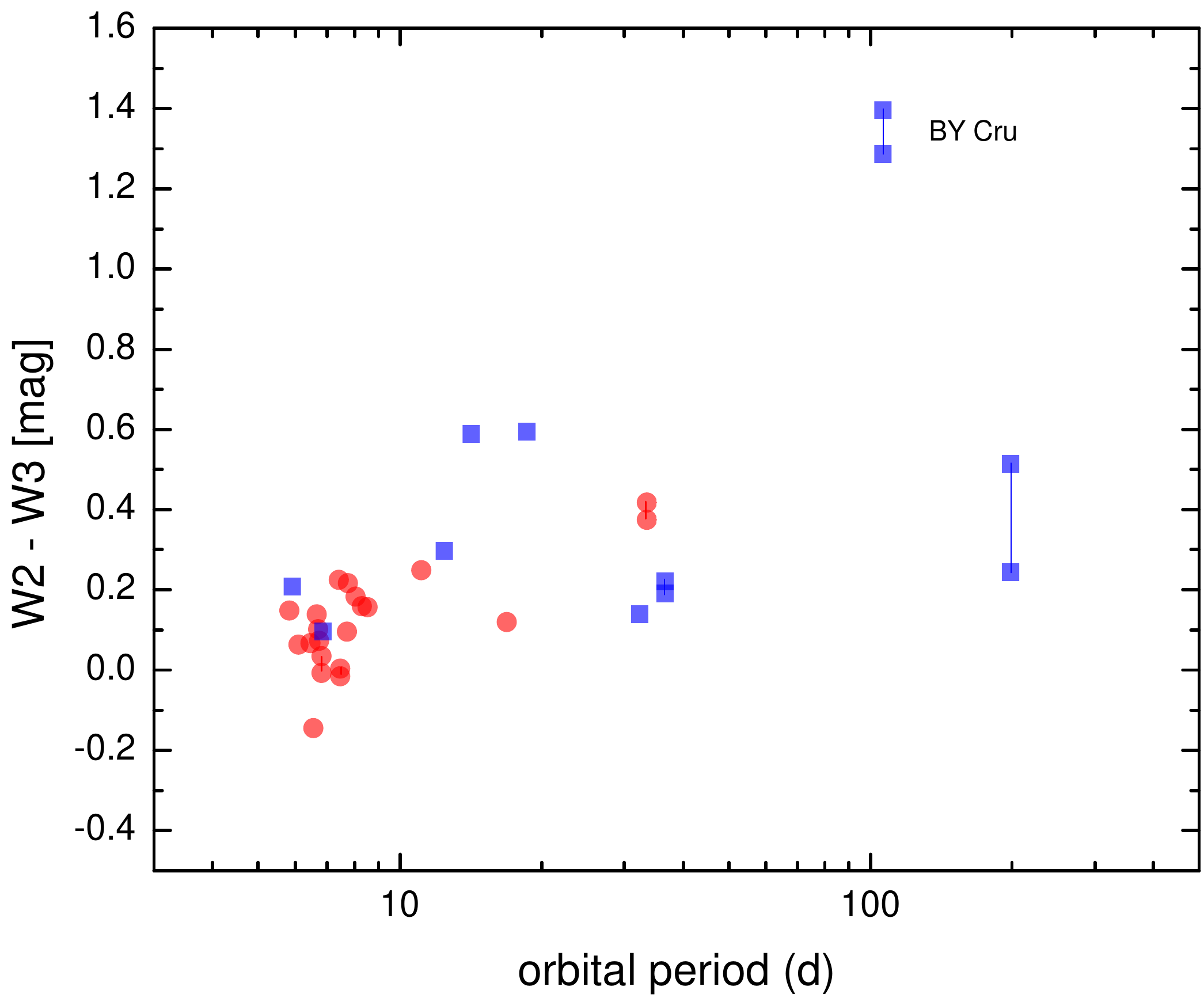}}
\caption{WISE colors versus orbital period for DPVs (red dots) and  W\,Serpentis stars (blue squares). 
Different epochs for a given star are represented by points joined by a line.}
 \label{fig9}
\end{figure}

\begin{figure*}
\scalebox{1}[1]{\includegraphics[angle=0,width=15cm]{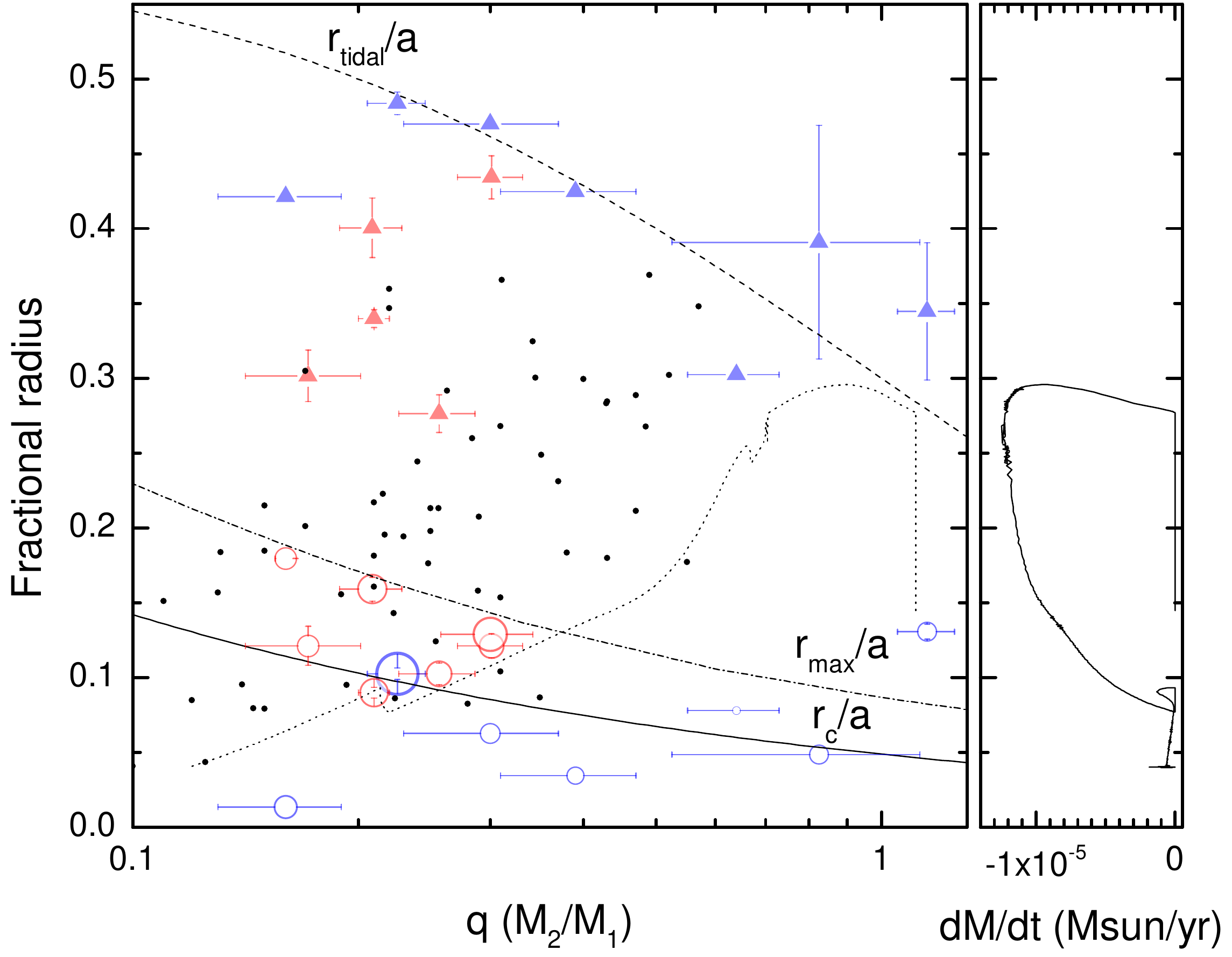}}
\caption{The relative radius for the primary ($R_{1}/a$; red circles for DPVs and blue circles for W\,Serpentids) and disc ($R_{d}/a$, triangles same color mark), according to the data given in Table 12.  Symbol size for stellar radius is proportional to the system total mass. When no $q$ error was available, we considered a representative error of 14\%. Below the
 circularization radius shown by the solid black line a disc should be formed and below  the dash-point a disc might be formed. 
 The tidal radius  indicates the maximum possible disc extension  (upper dashed line).
Semi-detached Algol primaries from Dervi{\c s}o{\v g}lu, Tout \& Ibano{\v g}lu  (2010) are also shown  as black points.
The track for $R_{1}/a$ for the best model of HD\,170582 is shown in the left panel and the corresponding mass transfer rate at the right panel (see text for details).}
 \label{fig10}
\end{figure*}

The third radius considered is the tidal radius (Paczynski 1977, Warner 1995):

\begin{equation}
\frac{r_{tidal}}{a} = \frac{0.6}{1+ q } ~~~~~\rm{for ~~0.03 < q < 1}.
\end{equation}

\noindent
This radius corresponds to the last non-intersecting orbit; it is assumed that a disc growing larger will start to
experience strong shear forces, hence it represents a limit for the disc size. 

From the inspection of Fig.\,10 and keeping in mind the low number of systems studied, we find that: (i)  most DPV primaries are found above the 
Lubow \& Shu (1975) critical radius  but below $r_{max}$,  (ii) W\,Serpentid gainers have radii normally below the critical radius, (iii) W\,Serpentid  discs usually reach the tidal radius and (iv) DPV discs are smaller than the tidal radius.
In this context, we note that a disc twice larger  than reported in Table\,12 was  found by Atwood-Stone et al. (2012) for AU\,Mon, by modeling the Balmer emission lines.
The difference was interpreted by the authors in terms of methods sensible to different disc regions;  whereas the fit of emission lines  is sensible to optically thin disc regions, the 
light curve model is sensible to the optically thick disc. Most disc radii in Table\,12 are derived from the analysis of light curves and represent the optically thick disc, rather than
optically thin Balmer emitting regions.

The finding that DPV gainers are larger than the critical radius  but smaller than $r_{max}$ is a surprising result, since  they should be candidates for hosting transient and variable discs (Peters 2001), however we have found that all DPVs have {\it stable} discs. The position of the DPV gainers in the diagram indicates they are impact systems.  However the impact of the stream is almost tangential, probably imparting most of the angular momentum to the star and
accelerating it more efficiently that a head-on impact.   The fact that DPVs possess a disc indicates that
the disc was formed in spite of the tangential impact. Theoretical work indicates that the impact should rapidly accelerate the star until critical rotation (e.g. Packet 1981). This fact suggests that after critical rotation
the extra supplied mass cannot be accumulated on the star, but starts forming an accretion disc in DPVs. The DPV characteristic of having pretty much stable orbital light curves, indicates that these discs, 
formed in a different way that discs of disc-systems  (those with $R_{1} < r_{c}$) ), are relatively stable.  It is possible that rapid rotation favors disc formation (see section 4.1.2). At present, it is not known why these discs are truncated to smaller radii than the tidal radius. The possibility that the disc is formed in earlier epochs, before reaching the tangential-impact condition, is explored at the end of Section 4.2.

The situation for W\,Serpentis stars is different since with the exception of UX\,Mon, most of them have primaries smaller than the critical radius, or of similar size.
Their discs might be formed in the usual way; the gas stream turns around the star, hits itself forming a hotspot and a ring that subsequently spreads forming an accretion disc. 
For W\,Serpentis stars the discs extend almost up to the tidal instability radius.
UX\,Mon is the only W\,Serpentid with $q > 1$  and contrary to all others W\,Serpentis systems, it has a primary much larger than the  critical radius; this system could have recently initiated the phase of rapid mass transfer, 
being still before the mass-ratio reversal, as suggested by Sudar et al. (2011). Therefore, the system is in a evolutionary stage different from all those considered here.

According to the above arguments, our study would benefit  from a comparison with the projected rotational velocities of the primaries.
However, the empirical determination of the gainer spin velocity in these systems is very difficult, since the presence
of the accretion disc introduces absorption/emission features contaminating the photospheric lines. For instance,
optical helium absorptions, potentially good indicators, are quite variable in shape and width during the whole orbital cycle.
Something similar happens for ultraviolet lines, especially in W\,Serpentis stars. 

 The light curve models described in Section 2.6 cannot discriminate between synchronous or critical rotation of the primary, since it is hidden by the accretion disc. We notice that the use of the equivalent radius of the primary, rather than the equatorial radius in DPVs, might underestimate the parameter $R_{1}/a$. For the extreme case of a critically rotating star the equatorial radius becomes 1.5 times the polar radius (Georgy et al. 2011). Since the equivalent radius is in between both radius, we expect an underestimation of  at most 25\%  in cases of critical rotation (Djura{\v s}evi{\'c} G., private communication).  This does not invalid the finding of a disc, that has been determined from the light-curve model, which is sensible to the size and shape of the disc rather than the radius of the primary. However, in cases of critical rotation, it should move some DPVs  slightly above the $r_{max}/a$ limit, i.e. into the region of impact systems. In this region no disc is expected, hence the finding of unexpected stable discs in DPVs remains valid.


\subsubsection{A comparison with semi-detached Algols}

Among candidate systems to form transient discs, i.e. those with $r_{c} < R_{1} < r_{max}$, we find a clear segregation between Algols and DPVs  when plotting the total mass and the mass of the primary for these systems (Fig.\,11). 
The important thing is that DPVs turns to be much more massive than Algols  therefore certain range of primary masses are required for triggering the DPV phenomenon among low $q$ systems.  Notoriously, the two W\,Serpentids in this sample are W\,Ser and $\beta$\,Lyr,  both in the extreme of the mass distribution,  again suggesting no link between DPVs and W\,Serpentids. A search in the literature reveals that
some of the Algols  with $r_{c} < R_{1} < r_{max}$ show discs (e.g. SW\,Cyg, Richards et al. 2014; RX\,Gem, Olson \& Etzel 2015) whereas others do not (e.g. TW\,And, Manzoori 2014; KO\,Aql, Soydugan et al. 2007).  This is consistent with the traditional view that they should show transient discs (Peters 2001). However, in this range DPVs have stable discs but are restricted to the ranges $r_{c} < R_{1} < r_{max}$;  $7 M_{\odot} < M_{1} < 10 M_{\odot}$ and $8 M_{\odot} < M_{total} < 13 M_{\odot}$. This range of parameters seems to be exclusive of the
DPV phenomenon.

\subsubsection{Are DPV gainers Be stars?}

Having determined luminosities, masses and temperatures for DPV gainers in Section 3.2, it is  now clear that they share physical characteristics,
including position in the HR diagram and  possibly rapid rotation, with Be stars (e.g. Zorec, Fr\'emat \& Cidale 2005). We would like to rise the still speculative question whether DPV gainers are subject to the same mechanism producing stochastic mass ejections in Be stars, which is hitherto unknown (Rivinius, Carciofi and Martayan 2013). If the mechanism is synchronized with the orbit of a close stellar companion, it could produce regular mass ejections, producing the regular long-cycle variability typical of a DPV. At least a fraction of Be stars showing
long-term quasi-periodic variability can share with DPVs physical mechanisms inducing activity. More studies are needed to explore this possibility. 

\subsection{An evolutionary link between DPVs and W\,Serpentids?}

The different regions occupied by W\,Serpentids and DPVs in Fig.\,10 suggest an evolutionary link between both types of variables.
In this Section we explore such a possibility considering the path followed by $R_{1}/a$, $q$ and the mass transfer rate $\dot{M}$ during binary evolution. For that, we assume that the evolutionary history of  DPVs and W\,Serpentis stars can be represented by the binary star evolutionary models
described in Section 2.7. Moreover, we take one of these models as representative of the general evolution of the mass ratio, $\dot{M}$  and primary fractional radius for the systems under study.

To illustrate the prediction of a model, let's consider HD\,170582 (Mennickent et al. in preparation). This model was found among the grid of binary star evolutionary track by Van Rensbergen et al. (2011)   following the method described by Mennickent et al. (2012a) and turned to be conservative with initial masses of 6  M$_{\odot}$ and 2.4 M$_{\odot}$ and initial orbital period of 1.0  day.
The theoretical path $R_{1}/a$ (Fig.\,10) indicates that the system evolves from the upper right part of the diagram into the lower left part, lowering its mass ratio mainly due to the mass transfer happening  during epochs of Roche-lobe overflow. 
The accompanying  right panel shows the mass transfer rate as a function of $R_{1}/a$ indicating  that the maximum mass transfer rate occurs before reaching the critical radius (around fractional radius 0.27 in our example). 

Accordingly, if W\,Serpentids are systems with larger $\dot{M}$ than DPVs, as suggested by their larger variability, larger infrared excess and orbital period changes, and both kinds of systems are evolutionarily connected, then we should expect to find them 
above DPVs in the $R_{1}-q$ diagram, since they are hypothetically earlier in the  evolutionary history (DPVs are found after the peak of mass transfer). Surprisingly, this is not the case, W\,Serpentis stars 
are found below DPVs.  It is then possible that both kinds of objects are not linked by evolution.

 In addition we note that W\,Serpentids  span a mass range much wider than DPVs. Taking into account objects like RS\,Cep ($M_{tot}$ = 3.2 \msun)
or SX\,Cas   ($M_{tot}$ = 6.6 \msun) it is hard to imagine that these object can be DPV precursors  since they cannot produce early B-type components (apparently a condition for a DPV) by mass exchange. In principle, they might be considered as the outcome of a previous DPV phase, if large systemic mass loss has happened in the system. However, to account for the primary mass of some W\,Serpentids,  the gainer  should loose mass, which seems highly unlikely. For the above reasons, we suspect that  the less massive W\,Serpentids are not evolutionarily connected with DPVs. However, it is hard to  be conclusive about the other more massive systems, for instance $\beta$ Lyrae.


 Finally in this section we notice that the possibility that the disc is formed before the tangential-impact condition is not possible for HD\,170582,  since the system comes
from a higher mass ratio and larger fractional primary radius, i.e. from a zone where no disc is possible, as shown by the $R_{1}/a$ track in Fig.\,10.

\begin{figure}
\scalebox{1}[1]{\includegraphics[angle=0,width=8cm]{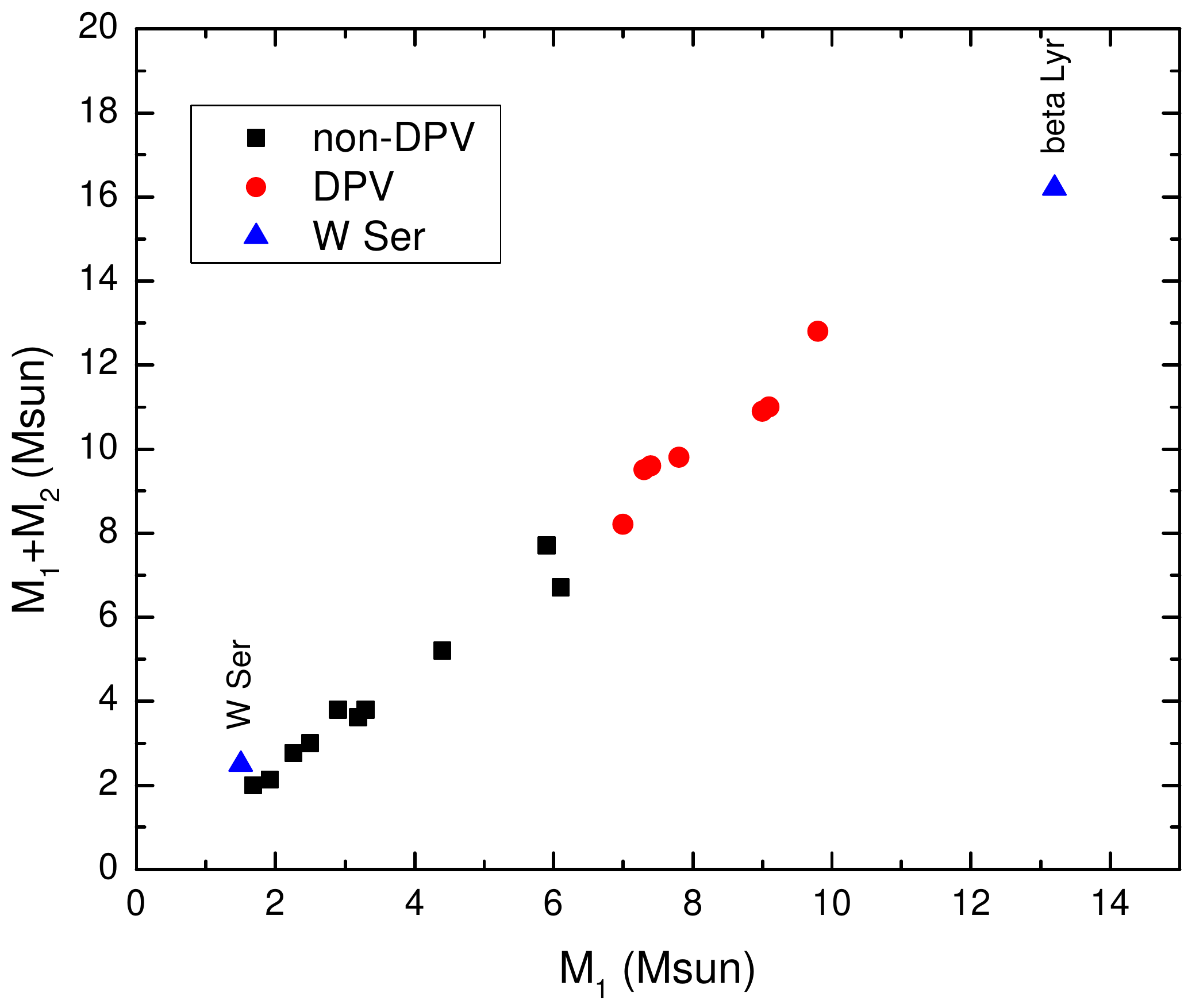}}
\caption{Primary mass and total mass of systems (Algols, DPVs and W\,Serpentids) with primaries with $r_{c} < R_{1} < r_{max}$. }
 \label{fig11}
\end{figure}

\section{Conclusions}

In this paper we have reported the discovery of 7 new Galactic Double Periodic Variables and listed some properties of all 21 known ones. 
We have also compared observational and physical parameters of W\,Serpentids and DPVs. Especially their infrared colors and properties of   accretion discs were also studied. 
 Whereas most of the 21 DPVs and 10 W\,Serpentids are included in the photometric study, only some of them have published physical data,
making our study of physical characteristics still based  on few cases, 7 DPVs and 7 W\,Serpentids. Keeping this low-number statistic restriction in mind, we arrive to the following conclusions: 

\begin{itemize}

\item Galactic DPVs show a  correlation between their long and orbital periods.  The long period is roughly 33 times the orbital one, but a range of period ratios are observed, between  27 and 39.
\item Among DPVs and W\,Serpentids, longer orbital period systems tend to show larger $H-K$ and $W2-W3$ colors. 
\item Contrary to previous reports (probably affected by difficulties in determining eclipse timings in light curves with variable shape), we find a constant orbital period for the W\,Serpentis system UX\,Mon, viz.\, $P_{o}$ = 5.90442 days. We notice that the linear ephemerides for the main minimum remains valid for at least 59 years.
\item In general, W\,Serpentids show larger infrared excess than DPVs. In both classes the excess, at least in some systems and excluding epochs of eclipses, is variable. We show that this can be understood in terms of variable amounts of circumstellar mass.
\item Among our sample, the system with the largest  $H-K$ and $W_{2}-W_{3}$  color excess  is BY\,Crucis.
\item DPV primaries are tangential-impact systems, i.e. they are slightly above the Lubow-Shu radius  and some of them might (if rotating critically) be barely inside the region of impact-systems.
Surprisingly, all of them show rather stable discs.
\item Discs of DPVs usually extend below the critical tidal radius. 
\item Discs of W\,Serpentids usually extend up to  the critical tidal radius.
\item Among impact systems, DPVs are those with primaries  corresponding to  slightly evolved B-type stars, with masses  in the range 7 $M_{\odot}$ $< M_{1} \lesssim $  10 $M_{\odot}$.   They should rotate rapidly sharing physical characteristics with Be stars, which are B-type fast rotators surrounded by disc-like envelopes whose mass ejection mechanism is still unknown.  In our sample the total mass of a DPV is in the range 8 $M_{\odot}$  $< M_{total} <$ 13 $M_{\odot}$. 

\end{itemize}

\section{Acknowledgments}

We thank an anonymous referee whose comments helped to improve a first version of this manuscript.  We also acknowledge Gojko Djura{\v s}evi{\'c} for useful discussions about this paper and Ahmet Dervi\c{s}o{\v g}lu for providing comparison data.
This publication makes use of VOSA, developed under the Spanish Virtual Observatory project supported from the Spanish MICINN through grant AyA2008-02156.
This research has made use of the SIMBAD database,
operated at CDS, Strasbourg, France.
This publication makes use of data products from the Wide-field Infrared
Survey Explorer, which is a joint project of the University of California, Los
Angeles, and the Jet Propulsion Laboratory/California Institute of Technology,
funded by the National Aeronautics and Space Administration. 
This research has made use of the NASA/ IPAC Infrared Science Archive, which
is operated by the Jet Propulsion Laboratory, California Institute of
Technology, under contract with the National Aeronautics and Space
Administration. This publication makes use of data products from the Two Micron All Sky Survey, which is a joint project of the University of Massachusetts and the Infrared Processing and Analysis Center/California Institute of Technology, funded by the National Aeronautics and Space Administration and the National Science Foundation.
R.E.M. acknowledges support by VRID-Enlace 214.016.001-1.0 and the BASAL Centro de Astrof{\'{i}}sica y Tecnolog{\'{i}}as Afines (CATA) PFB--06/2007. 
We acknowledge support from the Polish NCN
grant 2011/03/B/ST9/02667 to ZK.


\begin{table*}
\tiny
\centering
 \caption{The Galactic Double Periodic Variables ordered by increasing orbital period. The ratio between long and orbital periods is given, along with published spectral type and extreme visual magnitudes. References are given for the spectral type and for the paper announcing the long period (V360\,Lac) or DPV character. The single-wave (sw), double-wave (dw) or eclipsing (e) character of the orbital light curve is also given.} 
 \begin{tabular}{@{}lrrrrrrrrrrrrcc@{}}
\hline
Object &   Other name &   ASAS or NSVS ID &   RA & DEC &   Max $V$ &   Min $V$ &   $P_{\rm{o}}$ &   $P_{\rm{l}}$ &  $P_{\rm{l}}$/$P_{\rm{o}}$  &   Type &      SpT &   Ref.\,SpT  &  Ref.\,DPV\\
    &        &     & (2000)& (2000)&  (mag)  &  (mag)   &  (days) &(days) & & &  & & \\
\hline    
HD\,151582 &TYC\,7867-2398-1    &   ASAS\,J164954-3832.7 &   16 49 54.27 &-38 32 40.6 &   9.43 &   9.53     &   5.823 &   160 &   27.48 &   sw &     B3II/IIIe &   Houk (1982) &   3\\
DQ\,Vel &   TYC\,8175-333-1 &   ASAS\,J093034-5011.9 &   09 30 34.22& -50 11 54.1 &   10.73 &   11.69  &   6.0833 &   189 &   31.06 &   e &      B3V+A1III &   Barr\'{\i}a et al. (2013) &   1\\
BF\,Cir &   HD\,132461 &   ASAS\,J150232-6457.7 &   15 02 32.02 &-64 57 41.9 &   8.65 &   9.24  &   6.4592 &   219 &   33.87 &   dw &      B5V &   Houk \& Cowley (1975) &   3\\
GK\,Nor &   TYC\,8708-412-1 &   ASAS\,J153451-5824.0 &   15 34 50.91& -58 23 59.3 &   11.13 &   11.90 &   6.53971 &   225 &   34.44 &   e &-      & -   &   1\\
HD\,135938 &TYC\,8695-2281-1    &   ASAS\,J152008-5345.8 &   15 20 08.44 &-53 45 46.5 &   9.13 &   9.38     &   6.6477 &   231 &   34.78 &   dw &      B5/B6IVp &   Houk \& Cowley  (1975) &   1\\
HD\,50526 &TYC\,161-1014-1   &   ASAS\,J065402+0648.8 &   06 54 02.03 &+06 48 48.8 &   8.12 &   8.48     &   6.7007 &   192 &   28.61 &   dw &      B9 &   Ochsenbein (1980) &   1\\
V1001\,Cen &   HIP\,69978 &   ASAS\,J141910-5552.9 &   14 19 09.04 &-55 52 56.1 &   7.20 &   7.37     &   6.736 &   247 &   36.97 &   dw &     B4IV/V+OB: &   Dall et al. (2007) &   3\\
  NSV\,16849 & HD\,256413 &   ASAS\,J062402+1954.5 &   06 24 01.82 &+19 54 32.3 &   8.87 &   9.02     &   6.775 &   242 &   34.83 &   dw &      B5III &   Jaschek et al. (1964) &   3\\
HD\,90834 &TYC\,8613-1865-1    &   ASAS\,J102742-5917.3 &   10 27 41.61 &-59 17 04.9 &   9.08 &   9.42    &   6.815 &   231 &   33.90 &   dw &     B5/B6III/IV(e) &   Houk \& Cowley (1975) &   1\\
TYC\,5985-958-1 & GSC\,05985-00958   &   ASAS\,J074415-1758.8 &   07 44 15.30& -17 58 45.7 &   10.4 &   10.75     &   7.4054 &   229 &   30.87 &   dw &  -    &-    &   1\\
TYC\,8627-1591-1 & CPD-58\,3114   &   ASAS\,J110629-5848.3 &   11 06 29.07 &-58 48 18.8 &   8.77 &   8.94  &   7.462 &   268 &   35.92 &   dw &      B5 &Wallenquist (1931)    &   3\\
V393\,Sco &  HIP\,87191 &   ASAS\,J174848-3503.5 &   17 48 47.60& -35 03 25.6 &   7.39 &   8.31 &   7.71259 &   253 &   32.79 &   e &      B3III &   Houk (1982) &   1\\
V761\,Mon &   HIP\,36093 &   ASAS\,J072610-1032.9 &   07 26 09.54& -10 32 56.7 &   8.25 &   8.45 &   7.754 &   268 &   34.71 &   dw &      B5V+A: &   Houk (1999) &   3\\
CZ\,Cam &   HIP\,18593 &   NSVS\,512654 &   03 58 43.64 &+69 00 59.5 &   9.37 &   9.66     &   8.055 &   266 &   33.03 &   dw &      B5 &Heckmann (1975)    &   3\\
TYC\,5978-472-1 &CPD-21\,2186    &   ASAS\,J072641-2208.9 &   07 26 41.41 &-22 08 53.7 &   10.20 &   10.73     &   8.2958 &   312 &   37.61 &   dw &     B3V &   M\"unch (1952) &   1\\
LP\,Ara &   HD\,328568 &   ASAS\,J164002-4639.6 &   16 40 01.78 &-46 39 34.9 &   10.00 &   10.98     &   8.53295 &   273 &   31.99 &   e &     B8 &Nesterov et al. (1995)    &   1\\
V360\,Lac & HIP\,112778 &- &22 50 21.77 &+41 57 12.2 &5.88 & 5.99  &10.085 &322 &31.95 &dw &B3e+F9IV &Hill et al. (1997) &4 \\
AU\,Mon &   HIP\,33237 &   ASAS\,J065455-0122.5 &   06 54 54.71 &-01 22 32.8 &   8.20 &   9.16 &       11.11309 &   421 &   37.84 &   e &      B5e+G &   Desmet at al. (2010)&   1\\
HD\,170582 & TYC\,5703-2382-1    &   ASAS\,J183048-1447.5 &   18 30 47.53 &-14 47 27.8 &   9.60 &   9.86     &   16.871 &   537 &   31.83 &   dw &      A3III: &   Cannon \& Pickering (1922) &   1\\
V4142\,Sgr &   HD\,317151 &   ASAS\,J180745-2824.1 &   18 07 44.56& -28 24 04.3 &   10.85 &   12.70     &   30.636 &   1206 &   39.36 &   e &      A0 &   Nesterov et al. (1995) &   2\\
V495\,Cen &   CD-55\,4911 &   ASAS\,J130135-5605.5 &   13 01 34.81& -56 05 30.9 &   9.82 &   10.88     &   33.4873 &   1283 &   38.31 &   e &      Be+G2ep &   Skiff (2014) &   2\\
\hline
\end{tabular}
Ref.\,DPV:  (1) Mennickent et al. 2012a, (2) Mennickent \& Rosales 2014, (3) this paper, (4) Hill et al. 1997
\end{table*}

\begin{table*}
\centering
 \caption{Physical parameters for Galactic DPVs and  the LMC system OGLE\,05155332-6925581 (iDPV).   We give donor and gainer masses and radii ($M_{c}$, $M_{h}$ and $R_{c}$, $R_{h}$, respectively), the total system mass ($M_{tot}$), the mass ratio ($q$), the donor and gainer effective temperatures ($T_{c}$ and $T_{h}$, respectively), the logarithm of their luminosities in terms of the solar luminosity (log$(L_{c}/L_{\odot})$ and log$(L_{h}/L_{\odot})$, respectively), the system age, the hydrogen fraction of the stellar core for the donor ($X_{c}$) and the gainer ($X_{c}$) and the system mass transfer rate ($\dot{M}$ ).
 The last four parameters  in this table are from the models that best fit the observations. In all models the mass transfer rate ($\dot{M}$) is the same that the mass accretion rate. Errors can be found in the references. The systems are sorted by total mass.} 
 \begin{tabular}{@{}lrrrcccrlcccllcc@{}}
 \hline
DPV  &$M_{c}$ &$M_{h}$  &$M_{tot}$ &$q$ &$R_{c}$&$R_{h}$ &$T_{c}$ &$T_{h}$&log$(L_{c}/L_{\odot})$ &log$(L_{h}/L_{\odot})$  &$age$ &$X_{c}$ &$X_{h}$&$\dot{M}$ & {\rm ref}  \\
         &(M$_{\odot}$)    &(M$_{\odot}$)    &(M$_{\odot}$) & &(R$_{\odot}$) &(R$_{\odot}$) &(K) &(K) &  & &(yr) & & &(M$_{\odot}$ yr$^{-1}$)  &\\
 \hline
 LP\,Ara  &3.0 &9.8 &12.8 &0.30 &11.6 &5.3 &9500 &16400 &2.99 &3.26 & - & - &- &- &7 \\
  iDPV          &1.9 &9.1 &11.0 &0.21 &8.9 &5.6 &12900&25100&3.26&4.02&4.8E7 &0.0 &0.5 &3.1E-6 &1  \\
   HD\,170582 &1.9 &9.0 &10.9 &0.21 &15.6 &5.5 & 8000 &18000 &2.94 &3.46 &7.7E7 &0.0 &0.5 &1.6E-6&5 \\  
 V393\,Sco &2.0 &7.8 &9.8 &0.25 &9.4 &4.1 &7950&15850&2.46&3.06&7.0E7 &0.1 &0.6 &9.5E-9 & 2 \\
  V360\,Lac  &1.2 &7.4 &9.6 &0.16 &8.8 &7.6 &6000 &18000 &2.11 &3.54 &- &- &- &$<$3.2E-6 &6 \\
 DQ\,Vel   &2.2 &7.3 &9.5 &0.31 &8.4 &3.6 &9350&18600&2.66&3.14&7.4E7 &0.1 &0.5 &9.8E-9 & 3 \\
 AU\,Mon   &1.2 &7.0 &8.2 &0.17 &10.1 &5.1 &5750&15900&1.98&3.14&2.0E8 &0.0 &0.3 &7.6E-6 & 4,8 \\
   \hline 
\end{tabular}

References:  (1) Garrido et al. 2013, (2) Mennickent et al. 2012a, (3) Barr\'{\i}a et al. 2013, (4) Mennickent 2014, (5) Mennickent et al. 2015, (6) Linnell et al. 2006, (7) Mennickent et al. 2010, (8) Djura{\v s}evi{\'c} et al. 2010.
\end{table*}

\begin{table*}
\centering
 \caption{Physical parameters and additional data of W Ser stars.  Labels are as in Table\,3 and spectral types are from SIMBAD except for RY\,Per (Olson \& Plavec  1997), SX\,Cas (Andersen et al. 1988) and RS\,Cep (Olson \& Etzel 1995). The systems are sorted by total mass.} 
 \begin{tabular}{@{}lcrrrcccrrccccc@{}}
 \hline
system &$P_{o}$ &$M_{c}$ &$M_{h}$  &$M_{tot}$ &$q$ &$R_{c}$ &$R_{h}$ &$T_{c}$ &$T_{h}$&log$(L_{c}/L_{\odot})$&log$(L_{h}/L_{\odot})$  &$\dot{M}$/$\dot{P}$ &Sp & {\rm ref}  \\
        &(d)  &(M$_{\odot}$)    &(M$_{\odot}$)     &(M$_{\odot}$)  & &(R$_{\odot}$)  &(R$_{\odot}$)  & & &  &  &(M$_{\odot}$ yr$^{-1}$/(s yr$^{-1}$))   & &\\
 \hline
 $\beta$\,Lyr &12.94 &3.0  &13.2 &16.2 &0.22 &15.2 &6.0 &13200&30200&3.81&4.42  &1.6E-5/18.93 & B8II-IIIep &1  \\
 BY\,Cru &106.4 &1.7 &9.1 &10.8 &0.19 &52 &- & 11000 &- & 4.55 & - &-/-  &F0Ib-II  &7 \\
  W\,Cru &198.5 &1.2 & 7.8 &9.0 &0.16 &76 &4.0 &5500 &14000 & 3.68 & 2.74  &4.4E-8$-$1.3E-7/  &G2Iab  &4 \\
   RY\,Per &6.86 &1.6 &6.2 &7.8 &0.26 &8.10 &4.06 &6250 &18000 & 1.98 & 3.21  &-/-  &B4:V+F7:II-III &3 \\
 RX\,Cas &32.33 & 1.8 &5.8 &7.6 &0.30 &23.5 &2.5 &4400 &- &2.29   & -   &6E-6/19.86  &K1III+A5eIII&5 \\
 V367\,Cyg &18.6 &3.3 &4.0 &7.3 &0.82 &21.3 &2.9 &10400 &14800 &3.68 &2.56 &5-7E-5/ &B8peIa+F4III &11 \\
  UX\,Mon & 5.90 &3.9 &3.4 &7.3 &1.15 &9.80 &3.49 & 5990 &13000 & 2.04 & -  &-/- &A3  &2 \\
 SX\,Cas &36.6 &1.5 &5.1 & 6.6 &0.29 &23.5 &3.0 &4000&-  & 2.10 & -   &/-4.8  &A6(shell)+K3III &6 \\
RS\,Cep &12.42 &0.4 & 2.8 & 3.2 &0.14 &7.63 &2.65 & 4610 &9400&1.37 &1.69  &-/- &B9.7eV+G8III  &8 \\
W\,Ser &14.17 &1.0 &1.5  &2.5  &0.64 &1.00 &1.34 &- &- &-0.21 &0.45  &/14 &F5III  &9,10 \\

 \hline
\end{tabular}

References: (1) Mennickent \& Djura{\v s}evi{\'c} 2013. (2) Sudar et al. 2011, (3) Olson \& Plavec 1997, (4) Zo{\l}a 1996, (5) Andersen et al. 1989 , (6) Andersen et al. 1988, (7) Daems et al. 1997, (8) Olson \& Etzel 1995, (9) Budding et al.  2004, (10) Koch \& Guinan (1978), (11) Zo{\l}a \& Og{\l}oza 2001.
\end{table*}

\begin{table*}
\centering
 \caption{Errors for some of the physical parameters given in Tables\,3 and 4.} 
 \begin{tabular}{@{}lccccccrr@{}}
 \hline
system &$eM_{c}$ &$eM_{h}$  &$eM_{tot}$ &$eq$ &$eR_{c}$ &$eR_{h}$ &$eT_{c}$ &$eT_{h}$ \\
& (M$_{\odot}$)    &(M$_{\odot}$)     &(M$_{\odot}$)  & &(R$_{\odot}$)  &(R$_{\odot}$)  &($K$) &($K$) \\
 \hline
LP\,Ara &-&-&-&-&-&-&-&-\\
 iDPV     &0.20 &0.50&0.54&0.02&0.30&0.20&500&- \\
  HD\,170582 &0.10 &0.20&0.22&0.01&0.20&0.20&100&1500 \\
V393\,Sco &0.20 &0.50&0.54&0.03&0.30&0.20&300&500 \\
 V360\,Lac &0.05 &0.30&0.30&0.01&-&-&200&1000 \\
DQ\,Vel    &0.20 &0.30&0.36&0.03&0.20&0.20&100&500 \\
AU\,Mon   &0.20 &0.30&0.36&0.03&0.50&0.50&-&- \\
 $\beta$\,Lyr  &0.30 &0.30&0.40&0.02&0.20&0.20&-&- \\
 BY\,Cru  &- &-&-&-&-&-&-&- \\
  W\,Cru  &- &-&-&0.03&-&-&-&- \\
   RY\,Per  &0.10 &0.16&0.19&0.01&0.17&0.14&-&- \\
 RX\,Cas &0.40 &0.50 &0.60&0.07&1.20&-&-&- \\
  UX\,Mon  &0.29 &0.40&0.49&0.10&0.03&0.05&200&1000 \\
 SX\,Cas  &0.40 &0.40&0.60&0.08&1.30&0.40&300&- \\
RS\,Cep  &- &-&-&-&-&-&-&- \\
W\,Ser  &- &-&-&-&-&-&-&- \\
V367\,Cyg  &0.90 &0.50&1.00&0.30&2.00&-&-&- \\
 \hline
\end{tabular}
\end{table*}

\begin{table*}
\centering
\caption{Mean WISE magnitudes for Double Periodic Variables. The number of averaged measures and the variance of the mean are also given.  Epoch is given in julian day minus 2\,400\,000.} 
\begin{tabular}{@{}lccccccccccccc@{}}
 \hline
     Object   &       Epoch   &    NW1 & $W1 $&       $eW1 $ &  NW2  &  $W2 $  &    $eW2 $  & NW3 &   $W3 $  &  $ eW3 $  & NW4  &  $W4 $  &    $eW4 $ \\
\hline
V393\,Sco       &55273.9&    10  &6.3863  &0.0133&   11  &6.3843&  0.0094&    9&  6.2888&  0.0043&   10&  6.1727&  0.0282\\
 HD\,50526       &55285.5&    12  &7.6418  &0.0063&   13  &7.6082&  0.0071&   11&  7.5064&  0.0066&   11&  7.3702&  0.0696\\
  LP\,Ara       &55260.0&     3  &8.6070  &0.0058&    7  &8.6206&  0.0055&    3&  8.4633&  0.0489&    3& -&  -\\
 HD\,90834       &55208.6&     8  &8.5485  &0.0171&   10  &8.3591&  0.0131&   10&  8.3660&  0.0094&   11&  7.0731&  0.0584\\
 HD\,90834       &55382.7&    15  &8.3586  &0.0155&   14  &8.2610&  0.0163&   15&  8.2265&  0.0126&   17&  7.0882&  0.0810\\
  GK\,Nor       &55252.1&     4  &9.4543  &0.0054&    4  &9.4425&  0.0114&    5&  9.5876&  0.0962&    8&  7.7174&  0.0297\\
  CZ\,Cam       &55255.9&     8  &7.2654  &0.0061&    9  &7.1439&  0.0121&    9&  6.9609&  0.0112&    9&  6.6589&  0.0475\\
  AU\,Mon       &55286.4&    12  &7.4262  &0.0071&   14  &7.3321&  0.0072&   14&  7.0826&  0.0072&   14&  6.7645&  0.0667\\
V1001\,Cen      &55240.2&    12  &6.6881  &0.0084&   13  &6.6575&  0.0072&   13&  6.5844&  0.0035&   14&  6.4391&  0.0251\\
  DQ\,Vel       &55357.5&     3  &8.9337  &0.0088&    3  &8.8813&  0.0049&    4&  8.8182&  0.0319&   15&  7.5678&  0.0614\\
V495\,Cen       &55226.4&    11  &7.4510  &0.0101&   14  &7.3351&  0.0079&   12&  6.9176&  0.0063&   13&  6.3630&  0.0307\\
V495\,Cen       &55406.6&    12  &7.3406  &0.0041&   14  &7.1981&  0.0068&   11&  6.8238&  0.0039&   13&  6.2637&  0.0350\\
  BF\,Cir       &55251.1&    14  &7.3446  &0.0138&    9  &7.2959&  0.0171&    5&  7.2282&  0.0057&   11&  6.8606&  0.0289\\
TYC\,5978-472-1  &55298.8&    10  &8.9348  &0.0050&   11  &8.8644&  0.0047&   12&  8.7053&  0.0146&   12&  7.3874&  0.0826\\
 HD\,135938      &55248.7&    11  &8.1606  &0.0039&   12  &8.1215&  0.0053&   12&  7.9824&  0.0055&   11&  7.1285&  0.0653\\
 HD\,151582      &55260.6&    11  &7.9816  &0.0038&   11  &7.9063&  0.0093&    8&  7.7578&  0.0067&    9&  7.4073&  0.0601\\
 HD\,170582      &55283.1&    11  &7.7867  &0.0113&   11  &7.7081&  0.0088&   10&  7.5881&  0.0072&   10&  7.0755&  0.0549\\
TYC\,5985-958-1  &55302.8&    11  &9.7673  &0.0044&   12  &9.7267&  0.0054&   17&  9.5016&  0.0239&   13&  7.5242&  0.0697\\
TYC\,8627-1591-1 &55213.2&    15  &8.1965  &0.0048&   15  &8.1963&  0.0079&   16&  8.2111&  0.0090&   15&  7.4741&  0.0690\\
TYC\,8627-1591-1 &55388.6&    15  &8.1066  &0.0038&   15  &8.1043&  0.0042&   15&  8.1004&  0.0046&   16&  7.5813&  0.0441\\
 V761\,Mon      &55296.1&    13  &7.6085  &0.0060&   14  &7.5079&  0.0063&   14&  7.2911&  0.0044&   14&  6.9685&  0.0460\\
 \hline
\end{tabular}
\end{table*}


\begin{table*}
\centering
 \caption{Mean WISE magnitudes for W\,Serpentis stars. The number of averaged measures and the variance of the mean are also given.  Epoch is given in julian day minus 2\,400\,000.} 
 \begin{tabular}{@{}lccccccccccccc@{}}
 \hline
     Object   &       Epoch   &    NW1 &  $W1 $&      $ eW1 $ &  NW2  &  $W2 $  &    $eW2 $  & NW3 &   $W3 $  &   $eW3 $  & NW4  & $ W4 $  &   $ eW4 $ \\
   \hline
      W\,Ser      &55278.3&    10&  4.9480&  0.0049&    9&  4.5058&  0.0214&   11&  3.9167&  0.0056&   11&  3.3055&  0.0068\\
      W\,Cru      &55221.7&    16&  6.0051&  0.0223&   17&  5.7672&  0.0150&   13&  5.2522&  0.0067&   15&  4.8000&  0.0103\\
      W\,Cru      &55401.1&    15&  5.5384&  0.0164&   16&  5.1005&  0.0270&   16&  4.8562&  0.0106&   17&  4.3814&  0.0100\\
     SX\,Cas      &55213.6&    10&  6.0058&  0.0161&   10&  5.8752&  0.0128&   10&  5.6539&  0.0043&   11&  5.2675&  0.0102\\
     SX\,Cas      &55401.1&    16&  5.9231&  0.0037&   19&  5.8127&  0.0086&   17&  5.6208&  0.0027&   19&  5.2316&  0.0106\\
     RX\,Cas      &55247.1&     9&  5.3727&  0.0203&   10&  5.0776&  0.0280&    8&  4.9386&  0.0075&    8&  4.5903&  0.0147\\
     RY\,Per      &55235.7&     4&  7.7963&  0.0085&    4&  7.7548&  0.0080&    5&  7.6586&  0.0087&    8&  6.4769&  0.0346\\
     UX\,Mon      &55304.1&     4&  6.9363&  0.0059&    3&  6.9037&  0.0180&    3&  6.6957&  0.0224&    5&  6.3780&  0.0525\\
     RS\,Cep      &55266.3&    20&  8.4906&  0.0038&   19&  8.4586&  0.0056&   18&  8.1620&  0.0090&   19&  7.4905&  0.0413\\
     BY\,Cru      &55223.5&    15&  4.0338&  0.0063&   15&  2.8843&  0.0407&   13&  1.4884&  0.0129&   15&  0.7556&  0.0076\\
     BY\,Cru      &55403.3&    18&  4.1393&  0.0115&   18&  2.8846&  0.0484&   18&  1.5982&  0.0113&   16&  0.7499&  0.0064\\
   V367\,Cyg      &55339.1&    13&  4.7850&  0.0271&   17&  4.4098&  0.0320&   12&  3.8147&  0.0161&   12&  2.6889&  0.0087\\
   \hline
\end{tabular}
\end{table*}

\begin{table*}
\centering
 \caption{Mean WISE magnitudes for  the sample of Be stars studied by Howells et al.  (2001). The number of averaged magnitudes and the variance of the mean are also given.  Epoch is given in julian day minus 2\,400\,000.   } 
 \begin{tabular}{@{}lccccccccccccc@{}}
 \hline
     Object   &       Epoch   &    NW1 &  $W1 $&       $eW1  $&  NW2  &  $W2  $ &   $ eW2   $& NW3 &  $ W3  $ &  $ eW3  $ & NW4  &  $W4  $ & $   eW4 $ \\
   \hline
BD+05 3704&  55281.8& 11& 6.1935& 0.0141& 12& 6.2083& 0.0095& 12& 6.2894& 0.0030& 11& 6.2140& 0.0142\\
BD+17 4087&  55309.4& 13& 9.5734& 0.0044& 13& 9.5957& 0.0041& 12& 9.5604& 0.0180& 13& 7.4322& 0.0363\\
BD+20 4449&  55316.7& 12& 8.6632& 0.0033& 14& 8.7186& 0.0070& 13& 8.7003& 0.0075& 13& 6.8136& 0.0498\\
BD+21 4695&  55349.7& 15& 5.6497& 0.0117& 15& 5.2739& 0.0159&  4& 4.6550& 0.0064&  3& 3.7037& 0.0167\\
BD+23 1148&  55270.5&  9& 6.5306& 0.0098& 10& 6.5203& 0.0055& 10& 6.5312& 0.0065& 11& 6.3644& 0.0643\\
BD+27 797&  55265.1& 24& 7.2760& 0.0070& 24& 6.9868& 0.0050& 24& 6.2142& 0.0034& 24& 5.1681& 0.0170\\
BD+27 850&  55267.1& 11& 8.8902& 0.0051&  9& 8.9197& 0.0037& 11& 8.9821& 0.0346& 11& 7.4075& 0.0717\\
BD+27 3411&  55307.7& 19& 3.0899& 0.0942& 18& 2.4064& 0.1232& 14& 4.4551& 0.0031& 17& 3.6030& 0.0169\\
BD+28 3598&  55317.7& 17& 6.6895& 0.0239& 18& 6.8596& 0.0073& 17& 6.9085& 0.0048& 17& 6.7999& 0.0405\\
BD+29 3842&  55317.7& 18& 8.7521& 0.0050& 18& 8.7720& 0.0057& 16& 8.8202& 0.0095& 18& 7.6491& 0.0671\\
BD+29 4453&  55345.7& 14& 7.1104& 0.0104& 14& 6.7790& 0.0083& 12& 5.8019& 0.0030& 10& 4.2240& 0.0077\\
BD+30 3227&  55287.3& 19& 6.8265& 0.0081& 20& 6.8709& 0.0047& 19& 6.9060& 0.0039& 17& 6.7951& 0.0433\\
BD+31 4018&  55325.9& 15& 5.9698& 0.0080& 16& 5.5746& 0.0201& 16& 4.9844& 0.0038& 16& 4.3699& 0.0172\\
BD+36 3946&  55327.4& 18& 6.2966& 0.0081& 19& 5.9652& 0.0113& 17& 5.2534& 0.0030& 17& 4.4509& 0.0094\\
BD+37 675&  55234.6& 11& 5.9765& 0.0113& 11& 5.8060& 0.0147& 11& 5.3981& 0.0050& 11& 4.5746& 0.0071\\
BD+37 3856&  55328.7& 19& 9.3122& 0.0043& 19& 9.3522& 0.0073& 18& 8.8134& 0.0222& 18& 6.6343& 0.0386\\
BD+40 1213&  55261.6& 10& 6.5136& 0.0115& 12& 6.3053& 0.0058& 11& 5.7150& 0.0050& 12& 4.9535& 0.0099\\
BD+42 1376&  55267.1& 11& 7.0121& 0.0096& 12& 6.8702& 0.0065& 11& 6.0722& 0.0049& 12& 4.9454& 0.0086\\
BD+42 4538&  55375.6& 17& 7.4583& 0.0084& 15& 7.1718& 0.0019& 17& 6.1395& 0.0047& 17& 4.9861& 0.0110\\
BD+43 1048&  55257.1& 11& 8.3699& 0.0054& 11& 8.1750& 0.0089& 11& 7.5406& 0.0080& 11& 6.7283& 0.0598\\
BD+45 933&  55254.0& 11& 7.3565& 0.0086& 12& 7.3815& 0.0056& 12& 7.3992& 0.0096& 12& 6.4274& 0.0460\\
BD+46 275&  55217.6&  9& 4.3324& 0.0151&  9& 4.1378& 0.0256&  6& 4.3095& 0.0049&  7& 3.9483& 0.0123\\
BD+46 275&  55405.9& 15& 4.3300& 0.0068& 14& 4.1499& 0.0216& 13& 4.3540& 0.0044& 12& 4.0987& 0.0064\\
BD+47 183&  55214.4& 13& 3.9720& 0.0065& 13& 3.3056& 0.0351& 13& 2.8178& 0.0056& 13& 1.9732& 0.0086\\
BD+47 183&  55402.1& 14& 3.9733& 0.0098& 14& 3.4179& 0.0305& 12& 2.8296& 0.0040& 11& 1.9809& 0.0066\\
BD+47 857&  55243.9& 13& 3.6515& 0.0117& 13& 3.0137& 0.0305& 12& 2.4723& 0.0109& 11& 1.6250& 0.0079\\
BD+47 939&  55251.5& 13& 3.5626& 0.0133& 13& 2.9694& 0.0433&  5& 2.4942& 0.0245&  6& 1.6977& 0.0115\\
BD+47 3985&  55381.1& 35& 4.9382& 0.0093& 34& 4.4169& 0.0147& 27& 3.7499& 0.0028& 34& 2.8551& 0.0049\\
BD+49 614&  55231.3& 12& 7.5282& 0.0038& 13& 7.4612& 0.0061& 13& 6.8310& 0.0050& 13& 5.9392& 0.0201\\
BD+50 825&  55246.5& 20& 5.8037& 0.0092& 26& 5.7168& 0.0106& 21& 5.3420& 0.0065& 24& 4.5349& 0.0095\\
BD+50 3430&  55367.1& 19& 6.8457& 0.0075& 22& 6.7068& 0.0071& 20& 6.0545& 0.0049& 21& 5.1320& 0.0101\\
BD+51 3091&  55365.8& 21& 6.0315& 0.0137& 25& 5.9576& 0.0072& 23& 5.8773& 0.0030& 23& 5.2602& 0.0141\\
BD+55 552&  55233.8& 15& 7.8254& 0.0053& 15& 7.6070& 0.0047& 15& 7.0055& 0.0051& 16& 6.2826& 0.0237\\
BD+55 605&  55235.5& 14& 9.3514& 0.0064& 13& 9.3702& 0.0053& 15& 9.1095& 0.0214& 15& 7.3985& 0.0692\\
BD+55 2411&  55356.3& 31& 5.8959& 0.0078& 34& 5.8246& 0.0090& 32& 5.7749& 0.0031& 32& 5.2688& 0.0080\\
BD+56 473&  55234.8& 13& 8.1258& 0.0062& 14& 7.9419& 0.0063& 13& 7.1418& 0.0064& 14& 6.0912& 0.0231\\
BD+56 478&  55234.6& 28& 7.0385& 0.0080& 26& 6.7960& 0.0052& 28& 6.3310& 0.0028& 28& 5.7736& 0.0139\\
BD+56 484&  55234.7& 14& 7.8384& 0.0057& 14& 7.5403& 0.0079& 15& 6.7193& 0.0061& 15& 5.8327& 0.0253\\
BD+56 493&  55234.8& 30& 8.7677& 0.0036& 32& 8.6562& 0.0033& 32& 8.3571& 0.0064& 32& 7.4324& 0.0426\\
BD+56 511&  55235.0& 13& 7.9062& 0.0039& 15& 7.7807& 0.0049& 13& 7.2805& 0.0064& 15& 6.5809& 0.0577\\
BD+56 573&  55235.5& 14& 6.6435& 0.0132& 15& 6.1040& 0.0090& 15& 5.3573& 0.0057& 14& 4.8395& 0.0121\\
BD+57 681&  55241.7& 29& 7.3277& 0.0068& 29& 7.3072& 0.0057& 29& 7.3073& 0.0045& 28& 7.1215& 0.0526\\
BD+58 554&  55242.8& 14& 8.1644& 0.0040& 14& 8.0739& 0.0055& 14& 7.4791& 0.0052& 13& 6.5668& 0.0397\\
BD+58 2320&  55378.5& 25& 8.4337& 0.0077& 26& 8.2460& 0.0070& 25& 7.6126& 0.0062& 24& 7.1105& 0.0537\\
CD-22 13183&  55284.6& 22& 6.9400& 0.0045& 22& 6.8234& 0.0046& 22& 6.1277& 0.0040& 22& 5.2087& 0.0140\\
CD-28 14778&  55283.8& 11& 7.7360& 0.0034& 11& 7.4631& 0.0053& 11& 6.6760& 0.0032& 11& 5.7842& 0.0262\\
CD-27 11872&  55272.9& 18& 5.8685& 0.0222& 18& 5.6379& 0.0128& 16& 4.9731& 0.0027& 22& 3.9635& 0.0093\\
CD-27 16010&  55338.5& 17& 4.3160& 0.0086& 16& 4.0098& 0.0267& 15& 3.9422& 0.0061& 14& 3.2371& 0.0060\\
BD-12  5132&  55285.3& 11& 7.1323& 0.0174& 12& 6.8462& 0.0065& 10& 6.0340& 0.0058& 11& 5.2101& 0.0143\\
BD-19 5036&  55282.9& 12& 6.9371& 0.0088& 13& 6.9588& 0.0053& 13& 7.0665& 0.0055& 10& 7.1262& 0.0939\\
BD-02 5328&  55317.3&  9& 6.3130& 0.0075& 11& 6.2670& 0.0099& 11& 5.8419& 0.0036& 11& 4.9435& 0.0117\\
BD-01 3834&  55305.8&  8& 7.1134& 0.0076& 10& 6.7531& 0.0068&  9& 5.9466& 0.0036& 10& 5.1152& 0.0119\\
 \hline
\end{tabular}
\end{table*}

\begin{table*}
\centering
 \caption{Mean WISE magnitudes and spectral types for a sample of non-variable Hipparcos stars. The number of WISE averaged magnitudes and the variance of the mean are also given.} 
 \begin{tabular}{@{}rccccccccccccc @{}}
 \hline
    HIP        & NW1  &   $W1 $ &     $eW1  $&   NW2  &  $W2  $&    $ eW2 $ &    NW3 &   $W3 $  & $  eW3 $  &   NW4  &  $ W4 $  &   $ eW4  $    &   SpType\\
\hline
             57     &   15 & 6.1905 & 0.0142 &  15 & 6.2239 & 0.0105 &  15 & 6.2224 & 0.0042 &  15 & 6.1145 & 0.0237&    K2V\\
                  594     &   15 & 8.5200 & 0.0052 &  16 & 8.5381 & 0.0053 &  16 & 8.5099 & 0.0093 &   2 & 7.6230 & 0.0863&      A9V\\
        1585     &   13 & 8.2686 & 0.0049 &  12 & 8.2950 & 0.0058 &  13 & 8.2590 & 0.0118 &   3 & 7.6167 & 0.1158&    G2V\\
        1837     &   30 & 6.0640 & 0.0088 &  29 & 6.0820 & 0.0123 &  29 & 6.1111 & 0.0058 &  30 & 5.8806 & 0.0192&      K3/K4V\\
        2107     &   11 & 7.3050 & 0.0056 &  13 & 7.3325 & 0.0051 &  11 & 7.3356 & 0.0034 &  10 & 7.1229 & 0.0482&   G2/G3V\\
        2964     &   13 & 7.4816 & 0.0055 &  14 & 7.5044 & 0.0061 &  14 & 7.5005 & 0.0069 &   8 & 7.2986 & 0.0900& F0V\\
        3553     &   10 & 7.7460 & 0.0060 &  11 & 7.7802 & 0.0052 &  11 & 7.7622 & 0.0063 &   5 & 7.5634 & 0.1554&      G3V\\
        4537     &   14 & 7.3291 & 0.0107 &  14 & 7.3343 & 0.0037 &  13 & 7.3496 & 0.0042 &  12 & 7.1597 & 0.0750&      F5V\\
        5130     &   16 & 7.9609 & 0.0044 &  17 & 7.9916 & 0.0060 &  16 & 7.9589 & 0.0062 &   5 & 7.5070 & 0.0684&     F8V\\
        5462     &   38 & 7.0895 & 0.0048 &  40 & 7.1257 & 0.0035 &  40 & 7.0906 & 0.0029 &  40 & 6.7136 & 0.0254&     G5V\\
        5755     &   16 & 8.6863 & 0.0052 &  16 & 8.7153 & 0.0075 &  16 & 8.7153 & 0.0132 &   0 &    - &    -&   B8V\\
         5935     &   14 & 7.4049 & 0.0073 &  14 & 7.4471 & 0.0064 &  13 & 7.4411 & 0.0041 &  10 & 7.2664 & 0.0668&   F5V\\
          6169     &   15 & 7.5973 & 0.0072 &  14 & 7.5991 & 0.0048 &  15 & 7.5949 & 0.0069 &  12 & 7.2745 & 0.0723& F5V\\
        6664     &   13 & 8.2403 & 0.0058 &  14 & 8.2729 & 0.0072 &  13 & 8.2825 & 0.0094 &   3 & 7.5690 & 0.1673&    A8V\\
        6672     &   17 & 8.3842 & 0.0048 &  18 & 8.4011 & 0.0042 &  18 & 8.3796 & 0.0056 &   5 & 7.4748 & 0.1496&A3V\\
        8167     &   25 & 8.5554 & 0.0033 &  23 & 8.5894 & 0.0030 &  23 & 8.5380 & 0.0088 &   5 & 7.3620 & 0.1372&  F5V \\
        8373     &   13 & 8.2312 & 0.0058 &  14 & 8.2704 & 0.0058 &  13 & 8.2801 & 0.0091 &   2 & 7.7360 & 0.0438& G3V\\
        9499     &   19 & 6.4809 & 0.0086 &  21 & 6.5066 & 0.0074 &  19 & 6.4913 & 0.0049 &  20 & 6.4727 & 0.0291& F7/8V\\
       10360     &   11 & 7.7695 & 0.0053 &  11 & 7.7849 & 0.0075 &  10 & 7.7763 & 0.0070 &   7 & 7.4310 & 0.0783& F0V\\
       13089     &   20 & 7.7264 & 0.0042 &  19 & 7.7323 & 0.0047 &  21 & 7.7242 & 0.0044 &   9 & 7.4800 & 0.0656&  A9V\\
       13285     &   12 & 7.5673 & 0.0062 &  12 & 7.5822 & 0.0042 &  12 & 7.5495 & 0.0062 &   7 & 7.2986 & 0.0350& G0V\\
       15493     &   16 & 7.3433 & 0.0072 &  14 & 7.3863 & 0.0040 &  12 & 7.3515 & 0.0060 &  12 & 7.2620 & 0.0533&  K0V\\
       16465     &   11 & 8.3055 & 0.0055 &  11 & 8.3431 & 0.0052 &   9 & 8.3350 & 0.0076 &   4 & 7.5252 & 0.0817&     A3V\\
       16595     &   46 & 7.7703 & 0.0022 &  46 & 7.8192 & 0.0035 &  46 & 7.7668 & 0.0035 &  26 & 7.4697 & 0.0584&   G5/G6V\\
       17280     &   13 & 8.7922 & 0.0055 &  12 & 8.8215 & 0.0075 &  11 & 8.7695 & 0.0086 &   4 & 7.6240 & 0.0701&   B9V\\
       18719     &    9 & 6.8560 & 0.0087 &   9 & 6.9259 & 0.0101 &   9 & 6.9242 & 0.0027 &   9 & 6.8512 & 0.0679&    G4V\\
       19210     &   18 & 7.3509 & 0.0050 &  19 & 7.3670 & 0.0034 &  17 & 7.3604 & 0.0060 &  16 & 7.3072 & 0.0583&    F5V\\
       20019     &   18 & 6.4029 & 0.0085 &  14 & 6.3946 & 0.0182 &  13 & 6.3875 & 0.0074 &  21 & 6.2680 & 0.0297&    G8V\\
       20088     &   15 & 7.9141 & 0.0049 &  15 & 7.9505 & 0.0061 &  15 & 7.9247 & 0.0076 &   6 & 7.4895 & 0.0831&     G1V\\
       21159     &   10 & 7.9726 & 0.0080 &  10 & 7.9919 & 0.0086 &   8 & 8.0635 & 0.0104 &   3 & 7.5603 & 0.1435&    B5V\\    
       21472     &   14 & 7.7242 & 0.0055 &  15 & 7.7513 & 0.0073 &  13 & 7.7296 & 0.0070 &  11 & 7.4577 & 0.0588& A0V\\
       21498     &   13 & 8.1855 & 0.0037 &  14 & 8.1990 & 0.0044 &  12 & 8.1846 & 0.0068 &   6 & 7.6347 &   0.0766&   F0V\\
       22037     &   34 & 7.4060 & 0.0038 &  32 & 7.4141 & 0.0036 &  31 & 7.3943 & 0.0035 &  29 & 7.0341 & 0.0353&  F0V\\
       22295     &   30 & 6.8008 & 0.0063 &  30 & 6.8253 & 0.0035 &  31 & 6.8186 & 0.0039 &  32 & 6.5332 & 0.0199&   F7V\\
       23373     &   18 & 6.7974 & 0.0123 &  17 & 6.8061 & 0.0041 &  17 & 6.8210 & 0.0046 &  17 & 6.8294 & 0.0536&     F7V\\
       24922     &   14 & 8.4860 & 0.0055 &  14 & 8.5041 & 0.0085 &  13 & 8.4059 & 0.0100 &   7 & 7.3923 &  0.0672&    B8V\\
       25108     &   15 & 7.1920 & 0.0083 &  15 & 7.2328 & 0.0047 &  15 & 7.2418 & 0.0042 &  15 & 7.1533 & 0.0507& F5/F6V\\
       25321     &   14 & 6.3347 & 0.0088 &  15 & 6.3759 & 0.0085 &  15 & 6.3903 & 0.0038 &  14 & 6.2974 & 0.0291&   G8V\\
       27102     &   11 & 7.9664 & 0.0077 &  12 & 7.9914 & 0.0082 &  12 & 7.9475 & 0.0123 &   8 & 7.5802 & 0.0595& B9V\\
       27185     &   12 & 6.5821 & 0.0162 &  12 & 6.6044 & 0.0048 &  13 & 6.6986 & 0.0041 &  13 & 6.5483 & 0.0456&     G2V\\
       28247     &   23 & 7.7193 & 0.0030 &  25 & 7.7732 & 0.0046 &  23 & 7.7308 & 0.0045 &  17 & 7.4963 & 0.0756&     G6/G8V\\
       28923     &   17 & 8.9858 & 0.0046 &  18 & 9.0223 & 0.0048 &  17 & 9.0076 & 0.0131 &   3 & 7.7953 & 0.0729&    B8/B9V\\
       29644     &   15 & 6.7397 & 0.0092 &  15 & 6.8147 & 0.0060 &  13 & 6.8060 & 0.0077 &  12 & 6.7253 & 0.0309&     G8V\\
       30729     &   12 & 6.6851 & 0.0049 &  17 & 6.7240 & 0.0057 &  18 & 6.7273 & 0.0031 &  18 & 6.4988 & 0.0757&    G5V\\
       31036     &   12 & 7.2713 & 0.0103 &  10 & 7.2863 & 0.0057 &  11 & 7.3113 & 0.0058 &  13 & 7.1449 & 0.0786&    F2V\\
       32235     &   43 & 7.2850 & 0.0045 &  44 & 7.3144 & 0.0032 &  42 & 7.2705 & 0.0032 &  43 & 7.1295 &   0.0404&  G6V\\
       32326     &   10 & 7.2229 & 0.0088 &  10 & 7.2516 & 0.0073 &  10 & 7.2445 & 0.0051 &  10 & 7.2051 & 0.0629&   G2/G3V\\
       33208     &   14 & 9.2631 & 0.0040 &  17 & 9.3046 & 0.0058 &  14 & 9.2994 & 0.0113 &   2 & 7.6840 & 0.1520&   B3V\\
       34205     &   13 & 8.3539 & 0.0039 &  14 & 8.3788 & 0.0059 &  14 & 8.3548 & 0.0085 &   3 & 7.7450 &  0.1039  & A8V\\
       34230     &   11 & 6.8207 & 0.0139 &  12 & 6.8555 & 0.0061 &  11 & 6.7825 & 0.0059 &  11 & 6.7073 & 0.0617& G5V\\
       35117     &   15 & 9.1273 & 0.0052 &  15 & 9.1751 & 0.0052 &  13 & 9.1855 & 0.0124 &   2 & 7.0925 & 0.2493&   B9V\\
       36827     &   22 & 6.0054 & 0.0123 &  22 & 6.0195 & 0.0173 &  12 & 5.9892 & 0.0050 &  20 & 5.7438 & 0.0221&      K2V\\
       36944     &    9 & 8.9640 & 0.0051 &  10 & 9.0097 & 0.0097 &   9 & 9.0436 & 0.0106 &   0 &    - &    -&   B3V\\
       39687     &   20 & 8.7628 & 0.0055 &  19 & 8.8032 & 0.0044 &  19 & 8.8583 & 0.0093 &   1 & 7.8330 & 0.0000& B8V\\
       41587     &   20 & 7.4709 & 0.0056 &  20 & 7.4871 & 0.0038 &  18 & 7.4880 & 0.0042 &  15 & 7.4312 &  0.0951&  G3V\\
       43386     &   73 & 8.3713 & 0.0075 &  77 & 8.4064 & 0.0072 &  73 & 8.4143 & 0.0086 &  21 & 7.6716 & 0.0473&  B8V\\
       43562     &   14 & 8.1879 & 0.0031 &  14 & 8.2033 & 0.0033 &  14 & 8.1690 & 0.0082 &   7 & 7.5341 & 0.0490&    F8/G0V\\
       44101     &   15 & 7.2011 & 0.0071 &  15 & 7.2319 & 0.0060 &  15 & 7.2151 & 0.0060 &  15 & 7.1778 & 0.0543&     K1V\\
       44258     &   14 & 8.4189 & 0.0047 &  12 & 8.4395 & 0.0036 &  14 & 8.4435 & 0.0123 &   0 &    - &     -&    A1V\\    
       44526     &   20 & 6.2976 & 0.0046 &  26 & 6.3929 & 0.0050 &  25 & 6.3851 & 0.0037 &  26 & 6.2088 & 0.0172&      K2V\\
       45009     &   12 & 7.7062 & 0.0104 &  12 & 7.7256 & 0.0067 &   9 & 7.6978 & 0.0077 &   7 & 7.4709 & 0.0536& A0V\\
       \hline
\end{tabular}
\end{table*} 
     
     \begin{table*}
\centering
 \caption{Mean WISE magnitudes and spectral types for a sample of non-variable Hipparcos stars. The number of WISE averaged magnitudes and the variance of the mean are also given.} 
 \begin{tabular}{@{}rccccccccccccc @{}}
 \hline
    HIP        & NW1  &   $W1 $ &     $eW1  $&   NW2  &  $W2  $&    $ eW2 $ &    NW3 &   $W3 $  & $  eW3 $  &   NW4  &  $ W4 $  &   $ eW4  $    &   SpType\\
\hline  
       45857     &   16 & 7.9026 & 0.0057 &  16 & 7.9387 & 0.0068 &  15 & 7.9277 & 0.0094 &   8 & 7.5949 &   0.0700&   F6V\\
       47163     &   16 & 6.9721 & 0.0100 &  16 & 6.9927 & 0.0056 &  15 & 7.0032 & 0.0046 &  15 & 7.0963 & 0.0515&     G0V\\
       47743     &   14 & 7.8848 & 0.0029 &  12 & 7.9023 & 0.0051 &  15 & 7.9281 & 0.0104 &   8 & 7.3515 & 0.0909&   A2V\\
       48141     &   15 & 7.0227 & 0.0089 &  16 & 7.0600 & 0.0051 &  16 & 7.0543 & 0.0045 &  14 & 6.9986 & 0.0708&     G8V\\
       48627     &   16 & 7.6642 & 0.0052 &  15 & 7.6421 & 0.0066 &  14 & 7.6209 & 0.0061 &  12 & 7.4146 & 0.0811&   F0V\\
       49269     &   15 & 8.5885 & 0.0064 &  14 & 8.6186 & 0.0047 &  14 & 8.6057 & 0.0093 &   5 & 7.6614 & 0.0596&  B9V\\
       49636     &   13 & 9.1161 & 0.0054 &  12 & 9.1346 & 0.0064 &  12 & 9.1838 & 0.0130 &   1 & 7.7590 & 0.0000 & A0V\\
        49913     &   34 & 6.9441 & 0.0058 &  35 & 6.9949 & 0.0043 &  31 & 6.9919 & 0.0026 &  33 & 7.0311 & 0.0404& G1V\\
       50718     &   13 & 8.0399 & 0.0038 &  13 & 8.0797 & 0.0049 &  13 & 8.0400 & 0.0054 &   6 & 7.5937 & 0.0843&    G3V\\
       51534     &   16 & 7.7429 & 0.0055 &  18 & 7.7469 & 0.0060 &  15 & 7.7412 & 0.0104 &  10 & 7.4347 & 0.0839&      F0V\\
       52511     &   14 & 8.0116 & 0.0060 &  14 & 8.0154 & 0.0068 &  12 & 8.0171 & 0.0095 &   4 & 7.4727 & 0.0553& F2V\\
       52706     &   22 & 6.1241 & 0.0096 &  26 & 6.1397 & 0.0057 &  19 & 6.1563 & 0.0028 &  25 & 6.0932 & 0.0110&      G8V\\
       52716     &   20 & 6.9832 & 0.0073 &  19 & 7.0058 & 0.0061 &  18 & 7.0082 & 0.0047 &  19 & 7.0499 & 0.0509&    F7V\\
       52787     &   11 & 6.3790 & 0.0091 &  13 & 6.4629 & 0.0083 &  13 & 6.4546 & 0.0037 &  12 & 6.4153 & 0.0367& K0V\\
       53651     &   29 & 7.2672 & 0.0056 &  30 & 7.2957 & 0.0029 &  27 & 7.2985 & 0.0033 &  29 & 7.2931 & 0.0403&    F8/G0V\\
       55457     &   39 & 6.5985 & 0.0063 &  43 & 6.5951 & 0.0035 &  42 & 6.6297 & 0.0030 &  43 & 6.5545 & 0.0250&  F3V\\
       56264     &   18 & 8.6274 & 0.0044 &  17 & 8.6710 & 0.0043 &  17 & 8.6861 & 0.0118 &   2 & 7.8680 & 0.0559&     B6V\\
       57427     &   16 & 8.5301 & 0.0034 &  16 & 8.5724 & 0.0049 &  15 & 8.5067 & 0.0101 &   2 & 7.6910 & 0.1584& B9V\\
       59315     &   11 & 6.4349 & 0.0098 &  13 & 6.4614 & 0.0085 &  12 & 6.4708 & 0.0037 &  12 & 6.2839 & 0.0243&   G5V\\
       60679     &   14 & 7.3088 & 0.0116 &  14 & 7.3833 & 0.0073 &  13 & 7.3445 & 0.0051 &  11 & 7.2857 & 0.0705&    K0V\\
       63009     &   13 & 7.2270 & 0.0104 &  12 & 7.2648 & 0.0026 &  12 & 7.2548 & 0.0049 &  12 & 7.1032 &0.0834&     G5V\\
       63861     &   12 & 7.8967 & 0.0054 &  13 & 7.9258 & 0.0049 &  14 & 7.8940 & 0.0081 &   6 & 7.4772 & 0.0895&     G5V\\
       63862     &   11 & 6.6898 & 0.0071 &  12 & 6.7437 & 0.0057 &  13 & 6.7282 & 0.0044 &  13 & 6.7325 & 0.0593&  G5V\\
       64075     &   16 & 8.6369 & 0.0042 &  13 & 8.6577 & 0.0052 &  16 & 8.6921 & 0.0113 &   2 & 7.7045 & 0.0244& B9V\\
       65315     &   10 & 7.4352 & 0.0068 &  10 & 7.4614 & 0.0031 &  11 & 7.4535 & 0.0067 &   8 & 7.3440 &   0.1039&    G5V\\
        65186     &   25 & 8.2238 & 0.0043 &  24 & 8.2691 & 0.0028 &  25 & 8.2799 & 0.0084 &  21 & 7.0522 & 0.0481&  B8V\\
       67065     &   19 & 7.5635 & 0.0063 &  19 & 7.5832 & 0.0056 &  13 & 7.6118 & 0.0050 &  10 & 7.4691 & 0.0835&      F0V\\
       67191     &   12 & 7.1253 & 0.0132 &  11 & 7.1240 & 0.0076 &  11 & 7.1471 & 0.0077 &  10 & 7.0243 & 0.0679&  F2/F3V\\		
       67466     &   16 & 8.6491 & 0.0043 &  16 & 8.6614 & 0.0053 &  16 & 8.7285 & 0.0132 &   5 & 7.6896 & 0.0791&    B9V\\
       67907     &   16 & 6.9468 & 0.0131 &  16 & 7.0071 & 0.0056 &  17 & 7.0500 & 0.0058 &  15 & 6.9987 & 0.0432&    G0V\\
       69023     &   16 & 8.5389 & 0.0044 &  15 & 8.5537 & 0.0047 &  15 & 8.6191 & 0.0118 &   1 & 7.6620 & 0.0000&    F0V\\
       70477     &   14 & 8.2305 & 0.0042 &  12 & 8.2635 & 0.0035 &  14 & 8.3207 & 0.0079 &  13 & 7.4325 & 0.0598&     B4V\\
       71666     &   14 & 9.0378 & 0.0049 &  13 & 9.0673 & 0.0060 &  14 & 9.1400 & 0.0168 &   0 &    - &    -&      B3V\\
       72816     &   16 & 7.9609 & 0.0032 &  16 & 7.9915 & 0.0044 &  16 & 8.0381 & 0.0088 &   9 & 7.4046 & 0.0810& B5Vn\\
       74294     &   20 & 8.5834 & 0.0046 &  19 & 8.6008 & 0.0042 &  18 & 8.6158 & 0.0080 &   2 & 7.7030 &0.0849&     A2V\\
       75183     &   17 & 7.8358 & 0.0054 &  18 & 7.8781 & 0.0046 &  17 & 7.8541 & 0.0067 &   8 & 7.3822 & 0.0594&  G0V\\
       78299     &   15 & 7.2287 & 0.0072 &  14 & 7.2601 & 0.0054 &  13 & 7.2687 & 0.0053 &  13 & 7.1797 & 0.0540&   F3V\\
       80320     &   11 & 7.6194 & 0.0101 &  10 & 7.6345 & 0.0057 &  11 & 7.5530 & 0.0085 &   9 & 7.1217 & 0.0956&  G2/G3V\\
       80341     &   11 & 7.3750 & 0.0075 &  11 & 7.3951 & 0.0089 &  11 & 7.4414 & 0.0087 &   9 & 7.2601 & 0.0790&  F2V\\
       80535     &   11 & 7.1501 & 0.0094 &  10 & 7.1481 & 0.0062 &  10 & 7.1727 & 0.0076 &  10 & 7.1865 & 0.0663& G0V\\
       81334     &   12 & 7.4137 & 0.0077 &  12 & 7.4323 & 0.0080 &  12 & 7.4588 & 0.0098 &   9 & 7.1768 & 0.0989&  F2V\\
       82447     &   24 & 6.3267 & 0.0090 &  23 & 6.3477 & 0.0092 &  20 & 6.3794 & 0.0049 &  21 & 6.3005 & 0.0260&     G3/G5V\\
       82551     &   22 & 8.1692 & 0.0049 &   6 & 8.1875 & 0.0044 &  17 & 8.1355 & 0.0116 &   4 & 7.5727 &0.0974& A9V\\
       82577     &   16 & 8.3354 & 0.0055 &  16 & 8.3573 & 0.0051 &  15 & 8.3667 & 0.0078 &   6 & 7.7397 & 0.0667&    B9V\\
       84703     &   11 & 7.4937 & 0.0064 &  11 & 7.4916 & 0.0058 &  12 & 7.5183 & 0.0061 &   7 & 7.3931 & 0.0476& F0V\\
       85516     &   13 & 8.8232 & 0.0038 &  14 & 8.8399 & 0.0049 &  13 & 8.8098 & 0.0119 &   3 & 7.7613 & 0.0684&    F0V\\
       87292     &   15 & 6.9561 & 0.0109 &  16 & 7.0026 & 0.0052 &  15 & 6.9752 & 0.0055 &  15 & 6.9105 & 0.0602&      K3V\\
       88289     &   14 & 8.2190 & 0.0058 &  14 & 8.2473 & 0.0045 &  15 & 8.2463 & 0.0097 &   4 & 7.5765 & 0.1043&    A0V\\
       89907     &   10 & 8.0257 & 0.0031 &  11 & 8.0618 & 0.0049 &  12 & 8.0516 & 0.0079 &   7 & 7.4346 & 0.0833&   F0V\\
       93227     &   12 & 7.7947 & 0.0053 &  12 & 7.8174 & 0.0060 &  12 & 7.8254 & 0.0068 &   9 & 7.4704 & 0.0714& F0V\\
       94626     &   15 & 8.4136 & 0.0043 &  15 & 8.4280 & 0.0058 &  14 & 8.4084 & 0.0123 &   3 & 7.7793 & 0.0330&  F0/F2V\\
       95243     &   24 & 8.5514 & 0.0031 &  13 & 8.5670 & 0.0042 &  12 & 8.5198 & 0.0120 &   3 & 7.6757 & 0.1085&   A9V\\
       96303     &   26 & 7.0499 & 0.0067 &  28 & 7.0560 & 0.0058 &  28 & 7.0345 & 0.0048 &  26 & 6.8211 & 0.0336&    A5V \\
       99092     &   38 & 5.9763 & 0.0102 &  32 & 5.9977 & 0.0063 &  28 & 5.9656 & 0.0023 &  30 & 5.9360 & 0.0160&     K0V\\
       99412     &   19 & 8.3470 & 0.0045 &  19 & 8.3953 & 0.0059 &  17 & 8.4829 & 0.0099 &   0 &    - &    -&     B1V\\
      104580     &   20 & 7.5069 & 0.0063 &  19 & 7.5125 & 0.0033 &  19 & 7.5121 & 0.0047 &   7 & 7.2757 & 0.0640&    F5V\\
      104691     &   60 & 8.3692 & 0.0026 &  58 & 8.3792 & 0.0029 &  56 & 8.3345 & 0.0043 &  25 & 7.6545 & 0.0459&     F3/F5V\\
      105620     &    9 & 7.0793 & 0.0074 &   9 & 7.1117 & 0.0068 &   9 & 7.0972 & 0.0059 &  10 & 6.8834 &0.0593&    G6V\\
      107385     &   11 & 7.3453 & 0.0046 &  12 & 7.4013 & 0.0048 &  13 & 7.3595 & 0.0061 &  13 & 7.3830 & 0.0530& K2V\\
      108480     &   14 & 7.8047 & 0.0042 &  14 & 7.8454 & 0.0051 &  14 & 7.7986 & 0.0072 &  10 & 7.4160 & 0.0431& G2V\\
     \hline
\end{tabular}
\end{table*}

     \begin{table*}
\centering
 \caption{Mean WISE magnitudes and spectral types for a sample of non-variable Hipparcos stars. The number of WISE averaged magnitudes and the variance of the mean are also given.} 
 \begin{tabular}{@{}rccccccccccccc @{}}
 \hline
    HIP        & NW1  &   $W1 $ &     $eW1  $&   NW2  &  $W2  $&    $ eW2 $ &    NW3 &   $W3 $  & $  eW3 $  &   NW4  &  $ W4 $  &   $ eW4  $    &   SpType\\
\hline
      110659     &   19 & 7.1374 & 0.0079 &  20 & 7.1784 & 0.0059 &  22 & 7.1553 & 0.0039 &  20 & 6.9695 & 0.0364& G2V\\
        111252     &   33 & 7.6631 & 0.0093 &  35 & 7.6702 & 0.0073 &  25 & 7.6841 & 0.0062 &  20 & 7.5951 & 0.0445& A2V  \\
      112825     &   30 & 8.0623 & 0.0029 &  30 & 8.0977 & 0.0029 &  30 & 8.0462 & 0.0055 &  13 & 7.6444 & 0.0478&     F6V\\
      113071     &   11 & 8.5904 & 0.0069 &  11 & 8.6048 & 0.0056 &  11 & 8.6101 & 0.0174 &   4 & 7.3853 & 0.0555&     A1V\\
      113701     &   12 & 6.2458 & 0.0173 &  12 & 6.2460 & 0.0107 &  12 & 6.2789 & 0.0046 &  12 & 6.2347 & 0.0298&  K1V\\
      114198     &   10 & 7.4420 & 0.0108 &  10 & 7.4582 & 0.0069 &  10 & 7.4404 & 0.0090 &   8 & 7.3285 & 0.0512&  F0/F2V\\
      114879     &   14 & 7.3284 & 0.0065 &  15 & 7.3596 & 0.0048 &  15 & 7.3639 & 0.0037 &  14 & 7.3174 & 0.0639&   F0V\\
      114997     &   29 & 7.5807 & 0.0050 &  29 & 7.5979 & 0.0035 &  28 & 7.5829 & 0.0046 &  21 & 7.4086 & 0.0490&   F0V\\
      115034     &   31 & 7.5080 & 0.0037 &  32 & 7.5127 & 0.0047 &  32 & 7.5438 & 0.0052 &  24 & 7.3797 & 0.0494& A6V\\
      115694     &   16 & 7.7596 & 0.0051 &  16 & 7.7808 & 0.0031 &  16 & 7.7750 & 0.0069 &   9 & 7.4123 &  0.0651&  F6/F7V\\
      116122     &   26 & 6.8153 & 0.0064 &  26 & 6.8475 & 0.0029 &  27 & 6.8439 & 0.0036 &  28 & 6.7234 & 0.0336&    G2V\\
      116944     &   12 & 8.5752 & 0.0052 &  12 & 8.5886 & 0.0071 &  12 & 8.6220 & 0.0085 &   3 & 7.7160 & 0.1045& B7V\\
      117086     &   10 & 7.9788 & 0.0093 &   9 & 7.9999 & 0.0045 &   9 & 7.9633 & 0.0060 &   4 & 7.5338 & 0.0761&    F0V\\
      117596     &   19 & 6.4351 & 0.0195 &  19 & 6.4310 & 0.0071 &  18 & 6.4336 & 0.0045 &  17 & 6.2812 & 0.0238&     G3V\\
      118056     &   12 & 7.3900 & 0.0034 &  12 & 7.4242 & 0.0057 &  10 & 7.4119 & 0.0043 &  11 & 7.2281 & 0.0483&    F7V\\  
      \hline
\end{tabular}
\end{table*}

\begin{table*}
\centering
 \caption{Primary radius, disc radius, binary separation and their errors.  Stellar and orbital  data are from Tables 3, 4 and 5 and references therein. Disc data are from references listed in Tables 3, 4  and 5, except for RX\,Cas (\gojko 1993a), SX\,Cas (\gojko 1993b) and W\,Ser (Weiland et al. 1995).} 
 \begin{tabular}{@{}lcrrrrrrrrrr@{}}
 \hline
System& Type& $R_{1}$&$e(R_{1})$ & $R_{d}$& $e(R_{d})$ & $a$& $e(a)$& $R_{d}/a$&$e(R_{d}/a)$ & $R_{1}/a$&  $e(R_{1}/a)$ \\
& & (\rsun)&  (\rsun)&  (\rsun)&  (\rsun)&  (\rsun)&  (\rsun)& & & & \\
\hline
LP\,Ara& DPV& 5.30& -& -& -& 41.1& -& -& -& 0.13& - \\
HD\,170582& DPV& 5.50& 0.20& 20.80& 0.30& 61.2& 0.2& 0.34& 0.01& 0.09& 0.00\\
iDPV& DPV& 5.60& 0.20& 14.10& 0.50& 35.2& 0.5& 0.40& 0.02& 0.16& 0.01\\
V360\,Lac& DPV& 7.50& -& -& -& 41.8& -& -& -& 0.18& -\\
AU\,Mon& DPV& 5.10& 0.50& 12.70& 0.60& 42.1& 0.4& 0.30& 0.02& 0.12& 0.01\\
V393\,Sco& DPV& 3.60& 0.20& 9.70& 0.30& 35.1& 0.5& 0.28& 0.01& 0.10& 0.01\\
DQ\,Vel& DPV& 3.60& 0.20& 12.90& 0.30& 29.7& 0.3& 0.43& 0.01& 0.12& 0.01\\
W\,Cru& WSer& 4.00& -& 126.00& -& 299.0& -& 0.44& -& 0.01& -\\
RX\,Cas& WSer& 2.50& -& 18.80& -& 40.0& -& 0.47& -& 0.06& -\\
SX\,Cas& WSer& 3.00& 0.40& 37.00& -& 87.1& -& 0.42& -& 0.03& -\\
$\beta$\,Lyrae& WSer& 6.00& 0.20& 28.30& 0.30& 58.5& 0.3& 0.48& 0.01& 0.10& 0.00\\
UX\,Mon& WSer& 3.49& 0.05& 9.20& 1.00& 26.7& 0.7& 0.34& 0.05& 0.13& 0.01\\
W\,Ser& WSer& 0.97& -& 5.20& -& 17.2& -& 0.30& -& 0.06& -\\
V367\,Cyg & WSer&2.90 &- &23.30 &2.50 &59.60 &5.50 &0.39 &0.08& 0.05 &- \\
\hline
\end{tabular}
\end{table*}


\bsp 
\label{lastpage}

\newpage


\begin{thebibliography}{}




\bibitem[\protect\citeauthoryear{Andersen, Pavlovski, 
\& Piirola}{1989}]{1989A&A...215..272A} Andersen J., Pavlovski K., Piirola V., 1989, A\&A, 215, 272 


\bibitem[\protect\citeauthoryear{Andersen et 
al.}{1988}]{1988A&A...207...37A} Andersen J., Nordstrom B., Mayor M., Polidan R.~S., 1988, A\&A, 207, 37 

\bibitem[\protect\citeauthoryear{Arnold et al.}{1979}]{1979AcA....29..243A} 
Arnold C.~N., Montle R.~E., Stuhlinger T.~W., Hall D.~S., 1979, AcA, 29, 
243 

\bibitem[\protect\citeauthoryear{Atwood-Stone et 
al.}{2012}]{2012ApJ...760..134A} Atwood-Stone C., Miller B.~P., Richards 
M.~T., Budaj J., Peters G.~J., 2012, ApJ, 760, 134 






\bibitem[\protect\citeauthoryear{Barr{\'{\i}}a et al.}{2013}]{2013A&A...552A..63B} Barr{\'{\i}}a D., Mennickent R.~E., Schmidtobreick L., Djura{\v s}evi{\'c} G., Ko{\l}aczkowski Z., Michalska G., Vu{\v c}kovi{\'c} M., Niemczura E., 2013, A\&A, 552, A63 


\bibitem[\protect\citeauthoryear{Bessell, Castelli, 
\& Plez}{1998}]{1998A&A...333..231B} Bessell M.~S., Castelli F., Plez B., 1998, A\&A, 333, 231 

\bibitem[\protect\citeauthoryear{Budding et 
al.}{2004}]{2004A&A...417..263B} Budding E., Erdem A., {\c C}i{\c c}ek C., Bulut I., Soydugan F., Soydugan E., Baki{\c s} V., Demircan O., 2004, A\&A, 417, 263 

\bibitem[\protect\citeauthoryear{}{}]{} Bystrov N. F., Polojentsev D. D., Potter H. I., Yagudin L. I., Zallez R. F., Zelaya J. A., 1994, The final FOCAT-S star catalogue for southern hemisphere. Bulletin d'Information du Centre de Donnees Stellaires, Vol. 44, p. 3

\bibitem[\protect\citeauthoryear{Cannon 
\& Pickering}{1922}]{1922AnHar..97....1C} Cannon A.~J., Pickering E.~C., 1922, AnHar, 97, 1 

\bibitem[\protect\citeauthoryear{Dachs, Kiehling, 
\& Engels}{1988}]{1988A&A...194..167D} Dachs J., Kiehling R., Engels D., 1988, A\&A, 194, 167 



\bibitem[\protect\citeauthoryear{Daems, Waelkens, 
\& Mayor}{1997}]{1997A&A...317..823D} Daems K., Waelkens C., Mayor M., 1997, A\&A, 317, 823 


\bibitem[\protect\citeauthoryear{}{}]{}Dall T. H. et al., 2007, A\&A, 470, 1201

\bibitem[\protect\citeauthoryear{de Mink et 
al.}{2014}]{2014ApJ...782....7D} de Mink S.~E., Sana H., Langer N., Izzard 
R.~G., Schneider F.~R.~N., 2014, ApJ, 782, 7 





\bibitem[\protect\citeauthoryear{Dervi{\c s}o{\v g}lu, Tout, 
\& Ibano{\v g}lu}{2010}]{2010MNRAS.406.1071D} Dervi{\c s}o{\v g}lu A., Tout C.~A., Ibano{\v g}lu C., 2010, MNRAS, 406, 1071 

\bibitem[\protect\citeauthoryear{Deschamps et 
al.}{2015}]{2015A&A...577A..55D} Deschamps R., Braun K., Jorissen A., Siess L., Baes M., Camps P., 2015, A\&A, 577, A55 


\bibitem[\protect\citeauthoryear{}{}]{}Desmet M. et al., 2010, MNRAS, 401, 418


\bibitem[\protect\citeauthoryear{Djura{\v 
s}evi{\'c}}{1993}]{1993Ap&SS.206..129D} Djura{\v s}evi{\'c} G., 1993a, Ap\&SS, 206, 129 

\bibitem[\protect\citeauthoryear{Djurasevic}{1993}]{1993Ap&SS.208...85D} Djura{\v s}evi{\'c}, G., 1993b, Ap\&SS, 208, 85 


\bibitem[\protect\citeauthoryear{Djura{\v s}evi{\'c} et 
al.}{2010}]{2010MNRAS.409..329D} Djura{\v s}evi{\'c} G., Latkovi{\'c} O., 
Vince I., Cs{\'e}ki A., 2010, MNRAS, 409, 329 


\bibitem[\protect\citeauthoryear{Eggleton}{2006}]{2006epbm.book.....E} 
Eggleton P., 2006, Evolutionary processes in binary and multiple stars, Cambridge Astrophysical Series 40, Cambridge University Press.






\bibitem[\protect\citeauthoryear{Garrido et al.}{2013}]{2013MNRAS.428.1594G} Garrido H.~E., Mennickent R.~E., Djura{\v s}evi{\'c} G., Ko{\l}aczkowski Z., Niemczura E., Mennekens N., 2013, MNRAS, 428, 1594 



\bibitem[\protect\citeauthoryear{Georgy, Meynet, 
\& Maeder}{2011}]{2011A&A...527A..52G} Georgy C., Meynet G., Maeder A., 2011, A\&A, 527, A52 



\bibitem[\protect\citeauthoryear{Griffin}{2002}]{2002AJ....123..988G} 
Griffin R.~E., 2002, AJ, 123, 988 



\bibitem[\protect\citeauthoryear{}{}]{}Gudel, M. \& Elias, II, N. M. 1996, in Astronomical Society of the Pacific Conference Series, Vol. 93, Radio Emission from
the Stars and the Sun, ed. A. R. Taylor \& J. M. Paredes, 312




\bibitem[\protect\citeauthoryear{Halbedel}{1989}]{1989PASP..101..995H} 
Halbedel E.~M., 1989, PASP, 101, 995 

\bibitem[\protect\citeauthoryear{Harmanec et 
al.}{1996}]{1996A&A...312..879H} Harmanec P., et al., 1996, A\&A, 312, 879 




\bibitem[\protect\citeauthoryear{}{}]{}Heckmann O., 1975, AGK 3. Star catalogue of positions and proper motions
north of -2.5 deg. declination, Hamburg-Bergedorf: Hamburger Sternwarte,
edited by Dieckvoss W.

\bibitem[\protect\citeauthoryear{Hessman 
\& Hopp}{1990}]{1990A&A...228..387H} Hessman F.~V., Hopp U., 1990, A\&A, 228, 387 


\bibitem[\protect\citeauthoryear{}{}]{}Hill G., Harmanec P., Pavlovski K., Bozic H., Hadrava P., Koubsky P.,
Ziznovsky J., 1997, A\&A, 324, 965

\bibitem[\protect\citeauthoryear{}{}]{}Houk N., Cowley A. P., 1975, Dept. of Astronomy, Univ. of Michigan Ann
Arbor, Catalogue of two dimensional spectral types for the HD stars, Vol. 1


\bibitem[\protect\citeauthoryear{}{}]{}Houk N., 1982, Dept. of Astronomy, Univ. of Michigan Ann Arbor, Catalogue of
two-dimensional spectral types for the HD stars, Vol. 3


\bibitem[\protect\citeauthoryear{}{}]{}Houk N., Swift C., 1999, Dept. of Astronomy, Univ. of Michigan Ann Arbor,
Catalogue of two-dimensional spectral types for the HD stars, Vol. 5

\bibitem[\protect\citeauthoryear{Howells et 
al.}{2001}]{2001A&A...369...99H} Howells L., Steele I.~A., Porter J.~M., Etherton J., 2001, A\&A, 369, 99 

\bibitem[\protect\citeauthoryear{Hutchings 
\& van Heteren}{1981}]{1981PASP...93..626H} Hutchings J.~B., van Heteren J., 1981, PASP, 93, 626 



\bibitem[\protect\citeauthoryear{}{}]{}Jaschek C., Conde H., de Sierra A.C., 1964, Catalogue of Stellar Spectra
Classified in the Morgan-Keenan System, Serie Astronomica, No 2., La Plata:
Observatorio Astronomico de la Universidad de la Plata

\bibitem[\protect\citeauthoryear{Kalv}{1979}]{1979TarOT..58....3K} Kalv P., 
1979, TarOT, 58, 3 




\bibitem[\protect\citeauthoryear{Koch 
\& Guinan}{1978}]{1978IBVS.1483....1K} Koch R.~H., Guinan E.~F., 1978, IBVS, 1483, 1 

\bibitem[\protect\citeauthoryear{Kondo, McCluskey, 
\& Parsons}{1985}]{1985ApJ...295..580K} Kondo Y., McCluskey G.~E., Jr., Parsons S.~B., 1985, ApJ, 295, 580 

\bibitem[\protect\citeauthoryear{Kreiner 
\& Ziolkowski}{1978}]{1978AcA....28..497K} Kreiner J.~M., Ziolkowski J., 1978, AcA, 28, 497 








\bibitem[\protect\citeauthoryear{Linnell et 
al.}{2006}]{2006A&A...455.1037L} Linnell A.~P., et al., 2006, A\&A, 455, 1037 

\bibitem[\protect\citeauthoryear{Lubow 
\& Shu}{1975}]{1975ApJ...198..383L} Lubow S.~H., Shu F.~H., 1975, ApJ, 198, 383 


\bibitem[\protect\citeauthoryear{Lucy}{2005}]{2005A&A...439..663L} Lucy L.~B., 2005, A\&A, 439, 663 




\bibitem[\protect\citeauthoryear{Manzoori}{2014}]{2014AN....335.1064M} 
Manzoori D., 2014, AN, 335, 1064 

\bibitem[\protect\citeauthoryear{Mennickent et 
al.}{2003}]{2003A&A...399L..47M} Mennickent R.~E., Pietrzy{\'n}ski G., Diaz M., Gieren W., 2003, A\&A, 399, L47 

\bibitem[\protect\citeauthoryear{Mennickent et 
al.}{2008}]{2008MNRAS.389.1605M} Mennickent R.~E., Ko{\l}aczkowski Z., 
Michalska G., Pietrzy{\'n}ski G., Gallardo R., Cidale L., Granada A., 
Gieren W., 2008, MNRAS, 389, 1605 

\bibitem[\protect\citeauthoryear{Mennickent et 
al.}{2010}]{2010MNRAS.405.1947M} Mennickent R.~E., Ko{\l}aczkowski Z., 
Graczyk D., Ojeda J., 2010, MNRAS, 405, 1947 



\bibitem[\protect\citeauthoryear{Mennickent et 
al.}{2011}]{2011IAUS..272..527M} Mennickent R.~E., Graczyk D., 
Ko{\l}aczkowski Z., Michalska G., Barr{\'{\i}}a D., Niemczura E., 2011, 
IAUS, 272, 527 


\bibitem[\protect\citeauthoryear{Mennickent et 
al.}{2012}]{2012MNRAS.421..862M} Mennickent R.~E., Djura{\v s}evi{\'c} G., 
Ko{\l}aczkowski Z., Michalska G., 2012a, MNRAS, 421, 862 

\bibitem[\protect\citeauthoryear{Mennickent et 
al.}{2012}]{2012arXiv1205.6848M} Mennickent R.,~E., Ko{\l}aczkowski Z., Niemczura 
E., Diaz M., Cure M., Araya I., Peters G., 2012b, MNRAS, 427, 607 


\bibitem[\protect\citeauthoryear{Mennickent 
\& Djura{\v s}evi{\'c}}{2013}]{2013MNRAS.432..799M} Mennickent R.~E., Djura{\v s}evi{\'c} G., 2013, MNRAS, 432, 799 

\bibitem[\protect\citeauthoryear{Mennickent}{2014}]{2014PASP..126..821M} 
Mennickent R.~E., 2014, PASP, 126, 821 

\bibitem[\protect\citeauthoryear{Mennickent 
\& Rosales}{2014}]{2014IBVS.6116....1M} Mennickent R.~E., Rosales J., 2014, IBVS, 6116, 1 



\bibitem[\protect\citeauthoryear{Mennickent et 
al.}{2015}]{2015MNRAS.448.1137M} Mennickent R.~E., Djura{\v s}evi{\'c} G., 
Cabezas M., Cs{\'e}ki A., Rosales J.~G., Niemczura E., Araya I., Cur{\'e} 
M., 2015, MNRAS, 448, 1137 

\bibitem[\protect\citeauthoryear{}{}]{}M\"{u}nch L., 1952, Bolet\'{\i}n de los Observatorios de Tonantzintla y Tacubaya Vol.
1 Num. 2, p. 1

\bibitem[\protect\citeauthoryear{}{}]{}Nesterov V. V., Kuzmin A. V., Ashimbaeva N. T., Volchkov A. A., Roeser S.,
Bastian U., 1995, A\&AS, 110, 367, The Henry Draper Extension Charts: A
catalogue of accurate positions, proper motions, magnitudes and spectral
types of 86933 stars 

\bibitem[\protect\citeauthoryear{}{}]{}Ochsenbein F., 1980, Bull. Inf. CDS, 19, 74 

\bibitem[\protect\citeauthoryear{Olson 
\& Etzel}{1994}]{1994AJ....108..262O} Olson E.~C., Etzel P.~B., 1994, AJ, 108, 262 


\bibitem[\protect\citeauthoryear{Olson 
\& Etzel}{1995}]{1995AJ....110.2385O} Olson E.~C., Etzel P.~B., 1995, AJ, 110, 2385 

\bibitem[\protect\citeauthoryear{Olson 
\& Plavec}{1997}]{1997AJ....113..425O} Olson E.~C., Plavec M.~J., 1997, AJ, 113, 425 



\bibitem[\protect\citeauthoryear{Olson 
\& Etzel}{2015}]{2015AJ....149..125O} Olson E.~C., Etzel P.~B., 2015, AJ, 149, 125 





\bibitem[\protect\citeauthoryear{Packet}{1981}]{1981A&A...102...17P} Packet W., 1981, A\&A, 102, 17 



\bibitem[\protect\citeauthoryear{Paczynski}{1977}]{1977ApJ...216..822P} 
Paczynski B., 1977, ApJ, 216, 822 


\bibitem[\protect\citeauthoryear{}{}]{} Parihar P., Messina S., Bama P., Medhi B. J., Muneer S., Velu C., Ahmad A., 2009, MNRAS, 395, 593
 

\bibitem[\protect\citeauthoryear{Pawlak et al.}{2013}]{2013AcA....63..323P} 
Pawlak M., et al., 2013, AcA, 63, 323 

\bibitem[\protect\citeauthoryear{Perryman et 
al.}{1997}]{1997A&A...323L..49P} Perryman M.~A.~C., et al., 1997, A\&A, 323, L49 

\bibitem[\protect\citeauthoryear{Peters}{2001}]{2001ASSL..264...79P} Peters 
G.~J., 2001, ASSL, 264, 79 



\bibitem[\protect\citeauthoryear{}{}]{}Plavec, M., 1980a, "The impact of IUE on Binary Star Studies", U.C.L.A. Astronomy and Astrophysics Preprint No. 95.

\bibitem[\protect\citeauthoryear{}{}]{}Plavec, M., 1980b, "Mass loss from Interacting Close Binary Systems", U.C.L.A. Astronomy and Astrophysics Preprint No. 112.



\bibitem[\protect\citeauthoryear{Plavec}{1982}]{1982NASCP2338..526P} Plavec 
M.~J., 1982, NASCP, 2338, 526 



\bibitem[\protect\citeauthoryear{Plavec, Weiland, 
\& Koch}{1982}]{1982ApJ...256..206P} Plavec M.~J., Weiland J.~L., Koch R.~H., 1982, ApJ, 256, 206 


\bibitem[\protect\citeauthoryear{Plavec 
\& Dobias}{1983}]{1983ApJ...272..206P} Plavec M.~J., Dobias J.~J., 1983, ApJ, 272, 206 

\bibitem[\protect\citeauthoryear{Plavec}{1989}]{1989SSRv...50...95P} Plavec 
M.~J., 1989, SSRv, 50, 95 



\bibitem[\protect\citeauthoryear{Plavec}{1989}]{1989SSRv...50...95P} Plavec 
M.~J., 1989, SSRv, 50, 95 



\bibitem[\protect\citeauthoryear{}{}]{}Pojmanski G., 2002, Acta Astronomica, 52, 397









\bibitem[\protect\citeauthoryear{Poleski}{2010}]{}Poleski R., Soszy{\'n}ski I., Udalski A., 
Szyma{\'n}ski M.~K., Kubiak M., Pietrzy{\'n}ski G., Wyrzykowski {\L}., 
Ulaczyk K., 2010, AcA, 60, 179 


\bibitem[\protect\citeauthoryear{Pols et al.}{1998}]{1998MNRAS.298..525P} 
Pols O.~R., Schr{\"o}der K.-P., Hurley J.~R., Tout C.~A., Eggleton P.~P., 
1998, MNRAS, 298, 525 





\bibitem[\protect\citeauthoryear{Puss 
\& Leedjarv}{1997}]{1997BaltA...6..395P} Puss A., Leedjarv L., 1997, BaltA, 6, 395 


\bibitem[\protect\citeauthoryear{Pustylnik et 
al.}{2007}]{2007A&AT...26..339P} Pustylnik I., Kalv P., Harvig V., Aas T., 2007, A\&AT, 26, 339 

\bibitem[\protect\citeauthoryear{Quiroga et 
al.}{2002}]{2002A&A...387..139Q} Quiroga C., Miko{\l}ajewska J., Brandi E., Ferrer O., Garc{\'{\i}}a L., 2002, A\&A, 387, 139 

\bibitem[\protect\citeauthoryear{Richards et 
al.}{2014}]{2014ApJ...795..160R} Richards M.~T., Cocking A.~S., Fisher 
J.~G., Conover M.~J., 2014, ApJ, 795, 160 

\bibitem[\protect\citeauthoryear{Rivinius, Carciofi, 
\& Martayan}{2013}]{2013A&ARv..21...69R} Rivinius T., Carciofi A.~C., Martayan C., 2013, A\&ARv, 21, 69 






\bibitem[\protect\citeauthoryear{Sarna}{1993}]{1993MNRAS.262..534S} Sarna 
M.~J., 1993, MNRAS, 262, 534 



\bibitem[\protect\citeauthoryear{Simon}{1997}]{1997A&A...319..886S} Simon V., 1997, A\&A, 319, 886 


\bibitem[\protect\citeauthoryear{}{}]{}Skiff B. A., 2014, Catalogue of Stellar Spectral Classifications, VizieR
On-line Data Catalog: B/mk. Originally published in: Lowell Observatory
(October 2014)



\bibitem[\protect\citeauthoryear{Skrutskie et 
al.}{2006}]{2006AJ....131.1163S} Skrutskie M.~F., et al., 2006, AJ, 131, 
1163 

\bibitem[\protect\citeauthoryear{Soydugan et 
al.}{2007}]{2007MNRAS.379.1533S} Soydugan F., Frasca A., Soydugan E., 
Catalano S., Demircan O., Ibano{\v g}lu C., 2007, MNRAS, 379, 1533 



\bibitem[\protect\citeauthoryear{Stellingwerf}{1978}]{1978ApJ...224..953S} 
Stellingwerf R.~F., 1978, ApJ, 224, 953 


\bibitem[\protect\citeauthoryear{Sudar et 
al.}{2011}]{2011A&A...528A.146S} Sudar D., Harmanec P., Lehmann H., Yang S., Bo{\v z}i{\'c} H., Ru{\v z}djak D., 2011, A\&A, 528, AA146 



\bibitem[\protect\citeauthoryear{}{}]{} Szczygiel D. M., Socrates A., Paczynski B., Pojmanski G., Pilecki B., 2008, Acta Astronomica, 58, 405
 
 \bibitem[\protect\citeauthoryear{Tarasov, Berdyugina, 
\& Berdyugin}{1998}]{1998AstL...24..316T} Tarasov A.~E., Berdyugina S.~V., Berdyugin A.~V., 1998, AstL, 24, 316 

\bibitem[\protect\citeauthoryear{van Rensbergen et 
al.}{2008}]{2008yCat..34871129V} van Rensbergen W., De Greve J.~P., De 
Loore C., Mennekens N., 2008, yCat, 348, 71129 

\bibitem[\protect\citeauthoryear{Van Rensbergen et al.}{2011}]{2011A&A...528A..16V} Van Rensbergen W., de Greve J. P., Mennekens N., Jansen K., de Loore C., 2011, A\&A, 528, A16

\bibitem[\protect\citeauthoryear{}{}]{}Wallenquist A., 1931, Annalen v.d. Bosscha-Sterrenwacht, Vol. 3, 3.
gedeelte, Bandoeng: Nix, 1931, p. C3

\bibitem[\protect\citeauthoryear{Warner}{1995}]{1995CAS....28.....W} Warner 
B., 1995, Cambridge Astrophysics Series, 28  

\bibitem[\protect\citeauthoryear{Weiland et 
al.}{1995}]{1995ApJ...447..401W} Weiland J.~L., Shore S.~N., Beaver E.~A., 
Lyons R.~W., Rosenblatt E.~I., 1995, ApJ, 447, 401 



\bibitem[\protect\citeauthoryear{}{}]{} Whipple F. L., 1966, Smithsonian Astrophysical Observatory Star Catalog. Smithsonian Institution Press, Washington
 
 \bibitem[\protect\citeauthoryear{Wilson, Rafert, 
\& Markworth}{1984}]{1984IAPPP..16....1W} Wilson R.~E., Rafert J.~B., Markworth N.~L., 1984, IAPPP, 16, 1 

\bibitem[\protect\citeauthoryear{}{}]{}Wozniak P. R. et al., 2004, AJ, 127, 2436

\bibitem[\protect\citeauthoryear{Wright et al.}{2010}]{Wright10} Wright, E. L.,
Eisenhardt, P. R. M., Mainzer, A. K., et al. 2010, AJ, 140, 1868

\bibitem[\protect\citeauthoryear{Yoo}{2008}]{2008NewA...13..646Y} Yoo 
K.~H., 2008, NewA, 13, 646 

 \bibitem[\protect\citeauthoryear{Young 
\& Snyder}{1982}]{1982ApJ...262..269Y} Young A., Snyder J.~A., 1982, ApJ, 262, 269 

 

\bibitem[\protect\citeauthoryear{Zahn}{1977}]{1977A&A....57..383Z} Zahn J.~P., 1977, A\&A, 57, 383 


\bibitem[\protect\citeauthoryear{Zahn}{1975}]{1975A&A....41..329Z} Zahn J.~P., 1975, A\&A, 41, 329 



\bibitem[\protect\citeauthoryear{Zola}{1996}]{1996A&A...308..785Z} Zo{\l}a S., 1996, A\&A, 308, 785 



\bibitem[\protect\citeauthoryear{Zo{\l}a 
\& Og{\l}oza}{2001}]{2001A&A...368..932Z} Zo{\l}a S., Og{\l}oza W., 2001, A\&A, 368, 932 

\bibitem[\protect\citeauthoryear{Zorec, Fr{\'e}mat, 
\& Cidale}{2005}]{2005A&A...441..235Z} Zorec J., Fr{\'e}mat Y., Cidale L., 2005, A\&A, 441, 235 







\end{thebibliography}
\end{document}